\documentclass{vldb_tech}
\usepackage{graphicx}
\usepackage{balance}  
\usepackage{booktabs} 		
\usepackage{algorithm}
\usepackage{algpseudocode}
\usepackage{xcolor}			
\usepackage{lipsum} 		
\usepackage{xargs}
\usepackage{subfig}
\usepackage[colorinlistoftodos,prependcaption,textsize=tiny]{todonotes}
\usepackage{enumitem}

\algnewcommand\algorithmicforeach{\textbf{for each}}
\algdef{S}[FOR]{ForEach}[1]{\algorithmicforeach\ #1\ \algorithmicdo}

\algnewcommand\algorithmicinput{\textbf{Input:}}
\algnewcommand\Input{\item[\algorithmicinput]}

\algnewcommand\algorithmicoutput{\textbf{Output:}}
\algnewcommand\Output{\item[\algorithmicoutput]}

\newcommand{\toolName}{GALO} 

\newcommandx{\noteTodo}[2][1=]{\todo[linecolor=red,backgroundcolor=red!25,bordercolor=red,#1]{#2}}
\newcommandx{\noteChange}[2][1=]{\todo[linecolor=blue,backgroundcolor=blue!25,bordercolor=blue,#1]{#2}}
\newcommandx{\noteInfo}[2][1=]{\todo[linecolor=olive,backgroundcolor=olive!25,bordercolor=olive,#1]{#2}}

\newcommand{\col}[1]{\textcolor{black}{#1}}

\newcommand{\paddingTopFigure}{\vskip -0.30cm}
\newcommand{\paddingBottomFigure}{\vskip -0.50cm}
\newcommand{\paddingBottomFigureSmall}{\vskip -0.30cm}

\vldbTitle{GALO: Guided Automated Learning for query workloads re-Optimization}
\vldbAuthors{Guilherme Damasio, Vincent Corvinelli, Parke Godfrey, Piotr Mierzejewski, Alexandar Mihaylov, Jaroslaw Szlichta, and Calisto Zuzarte}

\vldbDOI{https://doi.org/10.14778/xxxxxxx.xxxxxxx}
\vldbVolume{12}
\vldbNumber{12}
\vldbYear{2019}
\makeatletter
\def\@copyrightspace{\relax}
\makeatother

\begin{document}


\vldbTitle{Guided automated learning for query \\workload re-optimization}

\title{Guided automated learning for query \\workload re-optimization}
\author{
\alignauthor
Guilherme Damasio{\small$^{^{\#}}$\small$^{^{\diamond}}$}, 
Vincent Corvinelli{\small$^{^{\$}}$}, 
Parke Godfrey{\small$^{^{*}}$\small$^{^{\diamond}}$},
Piotr Mierzejewski{\small$^{^{\$}}$},\\ 
Alex Mihaylov{\small$^{^{\#}}$\small$^{^{\diamond}}$},
Jaroslaw Szlichta{\small$^{^{\#}}$\small$^{^{\diamond}}$}, 
Calisto Zuzarte{\small$^{^{\$}}$}\\
        \affaddr{$^{\#}$\normalsize Ontario Tech University, Canada}
        \affaddr{$^{*}$\normalsize York University, Canada}\\
        \affaddr{$^{\diamond}$\normalsize IBM Centre for Advanced Studies, Canada},
        \affaddr{$^{\$}$\normalsize IBM Ltd, Canada}\\
        \affaddr{\normalsize 
        guilherme.fetterdamasio@uoit.ca, vcorvine@ca.ibm.com, godfrey@yorku.ca, piotrm@ca.ibm.com,\\ alexandar.mihaylov@uoit.ca, jarek@uoit.ca, calisto@ca.ibm.com}
}

\maketitle

\begin{abstract}

Query optimization is a hallmark of database systems 
When a SQL query runs more expensively than is viable or warranted, determination of the performance issues is usually performed manually in consultation with experts through the analysis of query's execution plan (QEP). However, this is an excessively time consuming, human error-prone, and costly process. GALO is a novel system that automates this process. The tool automatically learns recurring problem patterns in query plans over workloads in an offline learning phase, to build a knowledge base of plan-rewrite remedies. It then uses the knowledge base online to re-optimize queries 
often quite drastically.

GALO's knowledge base is built on RDF and SPARQL, W3C graph database standards, which is well suited for manipulating and querying over SQL query plans, which are graphs themselves. GALO acts as a third-tier of \col{re-optimization}, after query rewrite and cost-based optimization, as a \emph{query plan rewrite}. \col{For generality, the context of knowledge base problem patterns, including table and column names, is abstracted with canonical symbol labels. Since the knowledge base is not tied to the context of supplied QEPs, table and column names are matched automatically during the re-optimization phase. Thus, problem patterns learned over a particular query workload can be applied in other query workloads.} 
GALO's knowledge base is also an invaluable tool for database experts to debug query performance issues by tracking to known issues and solutions as well as refining the optimizer with new 
tuned techniques by the development team. We demonstrate an experimental study of the effectiveness of our techniques over synthetic TPC-DS and real IBM client query workloads.

\end{abstract}

\section{Introduction} \label{sec:introduction}
\subsection{Motivation}\label{sec:motivation}

As the complexity of queries, schemas, and database workloads
spiral ever upward,
the challenges in database systems have become severe.
SQL queries nowadays often are generated 
by middelware tools
instead of by SQL programmers~\cite{DBLP:conf/ismis/GryzWQZ08}.
Business-intelligence platforms such as IBM's Cognos
have enabled organizations to systematically scale data analysis
as never before.
These benefits, however, come with a price.
The generated SQL queries generated and run ``behind the scenes'' have
essentially no limit on their complexity,
often contain hundreds of algebraic operators
and span thousands lines of code.

Query optimization has long been a hallmark
of data warehouse systems,
which has truly enabled the data-analysis revolution~\cite{guravannavar2005optimizing,selinger1979access}.
However, there are cracks in the edifice;
the complexity of (automatically-generated) queries and workloads is 
outpacing what database systems can perform efficiently.
Database optimizers more often fail to pick best query plans.
	Research and development in query optimization is more vital today
	than it has ever \col{had} been,
	as people continue to address these new challenges~\cite{Ortiz:2018:LSR:3209889.3209890}.

Database vendors have made raw tools~\cite{Bruno:2009:4812427, Ziauddin:2008:OPC:1454159.1454175} available
to SQL programmers to troubleshoot performance problems
for given queries,
when the query optimizer fails to ``do the right thing''.
Oracle offers the keyword \emph{pragma} in its SQL,
which can be used by the programmer to override
decisions that the optimizer would make
concerning, for example, choice of join algorithms and join order.
(In truth, these pragma are suggestions to the optimizer.)
Likewise,
Microsoft SQL Server offers a similar mechanism via \emph{hints},
which are embedded in the SQL query.
IBM was reluctant to add a similar mechanism in DB2.
Pragma and hints can go stale over time
as \col{the database's statistics change},
yet they remain embedded in queries written in the past.
IBM took a different approach:
a \emph{guideline document} (written in XML) can be submitted
with a query to the optimizer.
Like pragma and hints,
the guidelines serve to sway the optimizer's choices in query planning.

The SQL programmer and database administrator (DBA) can
analyze the queries from a workload
with problematic performance, 
by profiling the query plans and execution traces,
to troubleshoot 
performance 
issues.
They then can override decisions in certain cases made by the database optimizer
for these problem queries
by using pragma, hints, and guidelines.

However, such performance debugging has become increasingly difficult
with very complex queries and workloads.
The causes of performance issues are, furthermore, often subtle.
More time than ever is
now spent by database system and optimization experts
at the major database-vendor companies
to help customers troubleshoot their workload performance problems.
This troubleshooting is often manual and painstaking. Automatic tools are needed for vendor experts,
SQL programmers and DBA's in the field
for this workload debugging. However, existing tools lack the ability to impose the proper structure for the execution plans of queries~\cite{agrawal2005database, Dageville:2004:AST:1316689.1316784, IBM:query.Workload.Tuner}. 

This workload debugging also has been \emph{ad hoc}.
The lessons learned from the fix for one problematic query
in one context---%
for a given database and system instance---%
are lost,
to be rediscovered by others later.
At IBM,
our recently developed OptImatch system
\cite{%
    Damasio2016ICDE,%
	Damasio2016EDBT%
}
has been a successful effort towards addressing this.
Experts feed problematic query-plan patterns and their resolutions
into an OptImatch \emph{knowledge base}.
Thus,
the expertise of problem resolutions
is systematically stored,
to be shared with
and queried by
others. Still, OptImatch knowledge base is built by experts manually by hand.
The job of troubleshooting new performance issues remains
exceedingly tedious and difficult.
The knowledge base helps, though,
experts to avoid re-solving previously solved issues,
and to find similar patterns that can help
with insights into the current issue.

\subsection{Goals}\label{sec:motivation}

In \toolName{},
we extend ambitiously on the original goals of OptImatch.
\toolName{}'s goals are threefold:

\begin{enumerate}
[nolistsep,leftmargin=*]
\item \label{GOAL/determine}
	\emph{automatic query problem determination};
\item \label{GOAL/re-opt}
	\emph{query re-optimization}; and
\item \label{GOAL/evolve}
	\emph{optimization evolution}.
\end{enumerate}

Goal \ref{GOAL/determine} is inherited from OptImatch.
\toolName{} significantly extends over OptImatch, however.
The knowledge base is \emph{automatically} ``learned''
rather than being manually
\col{constructed over the workloads}.
We use the RDF graph representation and the SPARQL language to impose the proper structure of execution plans for queries. \toolName{}'s architecture improves significantly 
on effectiveness and performance of the system,
as discussed in Section~\ref{section:System}.

Today's database optimizers are two stage:
a \emph{query-rewrite optimizer};
and
a \emph{cost-based optimizer}
\cite{Si1996}.
SQL offers many advantages for optimizing.
The relational algebra offers many opportunities
for re-ordering operations
as much is associative and commutative.
A query can be greatly rewritten
as long as the variant is semantically equivalent
to the original.
Query rewrite applies well-known, well-tested transformations
to an incoming query
to ``simplify'' it, so that
the resulting query plan will be more efficient.

The query-rewrite engine then passes the rewritten query
to the cost-based optimizer.
In cost-based optimization,
statistics of the database and system parameters
are used to make planning choices
based on cost estimations.
This generates a \emph{query execution plan} (QEP).
However,
there might be ``flaws'' in the chosen query plan.
Cost estimations may go awry.
Unusual characteristics in the data and the query
can circumvent the planning strategies as encoded in the optimizer.

In Goal \ref{GOAL/re-opt},
\toolName{} offers a \emph{third tier} of optimization,
\emph{query-plan} rewrite.
Rules from \toolName{}'s knowledge base can be applied
to the resulting query plan
that remove known performance trouble spots.
This is essentially an automation 
of the process done by hand
by SQL programmers
via pragma, hints, and guidelines.
Furthermore,
it applies all the acquired wisdom of performance debugging
via the knowledge base,
rather than ad hoc observations
from the SQL programmers. 
Also, it is applied at the time the query is to be run,
and not hard-coded into the SQL of the query itself
(as with pragma).

One way this query-plan rewrite could proceed
would be to ``patch'' the plan the cost-based optimizer produces
by applying the matched rewrites to it.
However,
this could result in incompatibilities in the overall plan.
Instead,
\toolName{} produces a guideline document with the chosen rewrites.
Then the query with the guidelines
is passed through the optimizer
(the query rewrite and the cost-based tiers) again to ensure that statistics and operators are updated over sub-portions of the generated plan.
This allows,
just as in the case of an SQL query with pragma, hints, or guidelines,
for the optimizer to generate a coherent query plan.
Not all guidelines may be honored,
as some may end up being incompatible within plan.
The cost-based optimizer will use the most profitable ones.
We call this \emph{re-optimization}.

Goal \ref{GOAL/evolve} is long-term.
\toolName{} can be utilized by the performance optimization team
to extract from the knowledge base those systemic issues
for the optimizer,
to learn and develop new rewrite rules for query rewrite
and new optimization techniques and refinements
for the cost-based optimizer.
And these improvements are not merely academic;
they arise directly from real-world workloads!
	\toolName{} has been well received within IBM,
	and is proving to be a valuable tool both
	in company support
	and in database optimizer development. 
\subsection{Real-world Example}\label{sec:example}

\begin{figure}[t]
\center
    \vskip -0.06cm
    \subfloat[Plan by IBM DB2.
		\label{fig:QEP_problem_1_a}]
	{
        \includegraphics[width=0.45\linewidth]
		{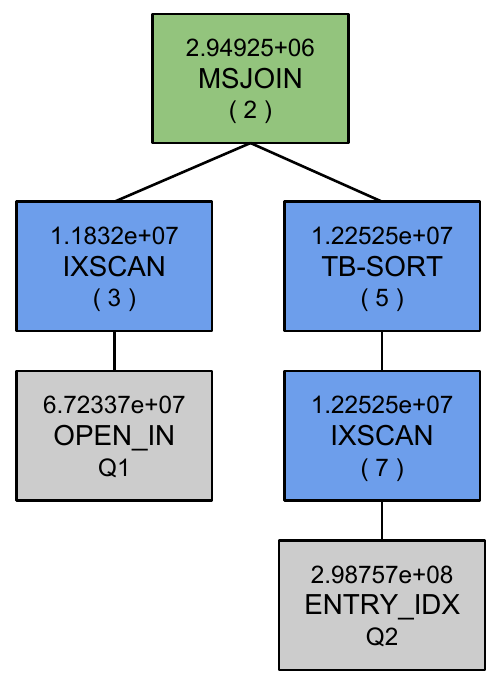}
    }
%
    \subfloat[Plan by \toolName{}.
		\label{fig:QEP_problem_1_b}]
	{
        \includegraphics[width=0.45\linewidth]
		{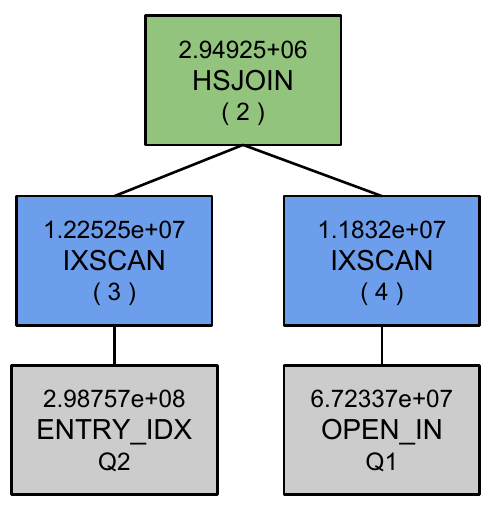}
    }
    \paddingTopFigure{}
    \caption{IBM client query with a problematic join.}
    \label{fig:QEP_problem_1}
   \paddingBottomFigureSmall{}
\end{figure}

Consider the portion of the tree of the query execution plan
chosen by IBM DB2 as ``optimal''
shown in Figure~\ref{fig:QEP_problem_1_a}.
The plan is comprised of a \emph{merge join} (\texttt{MSJOIN})
between the \texttt{OPEN\_IN} (\texttt{Q1})
and \texttt{ENTRY\_IDX} (\texttt{Q2}) tables.
Both are accessed via an \emph{index scan} (\texttt{IXSCAN}).
Each of the plan operators---%
e.g., \texttt{MSJOIN},
	\texttt{HSJOIN},\
	\texttt{TBSCAN},
	and
	\texttt{IXSCAN}---%
is referred in IBM DB2
as a \underline{lo}w \underline{le}vel
\underline{p}lan \underline{o}perator
(LOLEPOP).
The topmost decimal number of each LOLEPOP corresponds
to the optimizer's estimated cardinality,
and the integer in parentheses represents the operator ID.
For example,
in Figure~\ref{fig:QEP_problem_1_a},
the LOLEPOP with the \texttt{MSJOIN} has
an estimated cardinality of 2.94925e+06%
\footnote{%
	Cardinalities are integer, of course;
	however, since this is an estimation,
	a decimal floating point is used to represent it.
}
and an operator ID of 2.
For base tables,
the topmost decimal value corresponds to
the table's \emph{cardinality}
(the number of rows in the table),
and the value under the table name corresponds
to the \emph{instance} of the table.
For example,
the table \texttt{ENTRY\_IDX} has
an estimated cardinality of 2.98757e+08,
and a table instance value of \texttt{Q2}.


This pattern is an example of a real-life under-performing
query from one of the IBM customers.
The performance issue here hinges on the optimizer's join choice:
the \texttt{MSJOIN} reads the table \texttt{ENTRY\_IDX}
through an \texttt{IXSCAN} (\#7),
and then performs a sort
that is read by a table scan, \texttt{TB-SORT} (\#5).
The size of data into the sort
and the number of pages that spill to the disk at runtime are large.

The chosen \emph{fix} shown
in Figure \ref{fig:QEP_problem_1_b}
changes the type of the join and the join order.
This ``changes'' the \texttt{MSJOIN}
to a \emph{hash join} (\texttt{HSJOIN}),
and swaps the \emph{outer} and \emph{inner} tables as input
to the join
from \texttt{OPEN\_IN} on the left
and \texttt{ENTRY\_IDX} on the right
to \texttt{ENTRY\_IDX} on the left
and \texttt{OPEN\_IN} on the right.
While the \texttt{HSJOIN} spills pages into the disk at runtime too,
the amount is significantly smaller.

This plan rewrite reduced
the query runtime from nine hours to just five minutes! 
However, as it is evident from the rewrite in Figure \ref{fig:QEP_problem_1}, 
the steps required to fix existing performance issues are
not always intuitive or simple.
IBM experts report that even more complex patterns / rewrites exist
that are more daunting,
which can take days to be found and resolved.

\subsection{Contributions}\label{sec:contributions}

We developed the \toolName{} system,
which can serve as a third tier of optimization
by rewriting problematic portions of query plans
to result often in dramatically increased performance. Our main contributions are as follows.

\begin{enumerate}[nolistsep,leftmargin=*]
\item \textbf{The Knowledge Base.}
	First is \toolName{}'s \emph{knowledge base},
	an innovative and powerful representation
	for storing, manipulating, and querying SQL query-plan patterns.
%
	\toolName{}'s \emph{transformation engine} is responsible
	for mapping SQL queries and query plans
	to the knowledge base's RDF format,
	and to the SPARQL queries used to query the knowledge base.

\item \textbf{The Learning Engine.}
	Second is the \emph{learning engine},
	which is used \emph{offline}
	to populate the knowledge base.
	It analyzes large and complex SQL queries in the workload,
	and segments them into sub-queries.
	\col{The set of sub-queries is then broadened;
	for each sub-query,
	the values of the query's predicates are varied to result
	in different \emph{reduction factors}
	(and, hence, result cardinalities).}
	Then, for each sub-query from this broadened set,
	the query plan that the optimizer produces
	is compared against competing plans
	found using DB2's Random Plan Generator.
	Whenever a competing plan is found
	that performs significantly better than the optimizer's,
	it flags the pair as a potential \emph{rewrite}
	(a problematic plan pattern and its guideline solution).
	For \emph{abstraction},
	the table and column names in the plans
	are replaced by canonical symbol labels.

\item \textbf{The Matching Engine.}
	Third is the \emph{matching engine},
	which is employed \emph{online}
	to re-optimize the query plans of incoming queries
	by querying the knowledge base (\col{via SPARQL queries})
	to find matching plan rewrites.
	\col{Since the knowledge base's rewrites are abstracted (with canonical symbol labels for tables and attributes),
	a query with the similar sub-structure and characteristics
	can match a problem pattern of a rewrite
	that had been discovered
	during learning over a different query,
	even from a different query workload.
	The SPARQL queries generated for matching into the knowledge base
    support naturally this abstraction.
    SPARQL node-binding variables match to the canonical names.}
\end{enumerate}

\begin{enumerate}[nolistsep,leftmargin=*]
\setcounter{enumi}{3}
\item \textbf{Experiments.}
	\begin{enumerate}[nolistsep,leftmargin=*]
	\item
		We demonstrate experimentally dramatic query performance improvement and scalability of our solution over the TPC-DS benchmark and real-world IBM customer query workloads.
		\col{We also show that problem patterns learned over one query workload
		are re-used when re-optimizing queries in other workloads.}


	\item
		We quantify the benefits of our automatic approach
		against manual diagnosis.
		A collaborative study illustrates
		that the system is able to perform more effectively
		than IBM experts by providing more optimized solutions
		for problematic queries,
		while saving a significant amount
		of time to analyze QEP's.

	\end{enumerate}

\end{enumerate}

In Section \ref{section:System},
we overview \toolName{}'s architecture .
In Section~\ref{section:Modules},
we describe in detail the three primary components as discussed above,
the knowledge base, the learning engine,
and the matching engine.
In Section ~\ref{section:Experiments},
we provide a comprehensive experimental evaluation.
In Section~\ref{related-work},
we discuss the related work and we conclude in 
Section~\ref{conclusions}.
\section{\toolName{} System Overview} \label{section:System}

\toolName{} is an automated system to improve 
SQL \col{workload} performance.
We consider a \emph{workload} here to be
a populated database with a requisite schema
and a collection of SQL queries
that are periodically executed
on a given database system instance.
\toolName{} \emph{profiles} the \col{workloads} \emph{offline}
to construct a \emph{knowledge base}
which captures performance issues
(that have resolutions)
from the queries in the \col{workloads}.
When the workload is executed
(e.g., periodically in a data warehouse),
\toolName{} acts as a \emph{third stage}
of re-optimization
by applying \emph{rewrites} from the knowledge base (KB)
to the query plans \emph{online} to improve performance.

\toolName{} extends upon our OptImatch system
\cite{Damasio2016ICDE,Damasio2016EDBT}.
As the ``second generation'' of OptImatch,
\toolName{} is used within IBM
to populate automatically a general knowledge base
that tracks query plan issues
(as opposed to building such manually, as was done in OptImatch).
The knowledge base is used 
to resolve customers' workload performance issues.
\toolName{} is also being used as a resource
for the IBM DB2 engine team
for evolving the DB2 optimizer.
Rewrites in the knowledge base (improved query plans)
can be extracted and generalized
to inform the optimizer team,
where optimization rules should be added and refined.
%

\begin{figure}[t]
\center
     \includegraphics[scale=0.50]{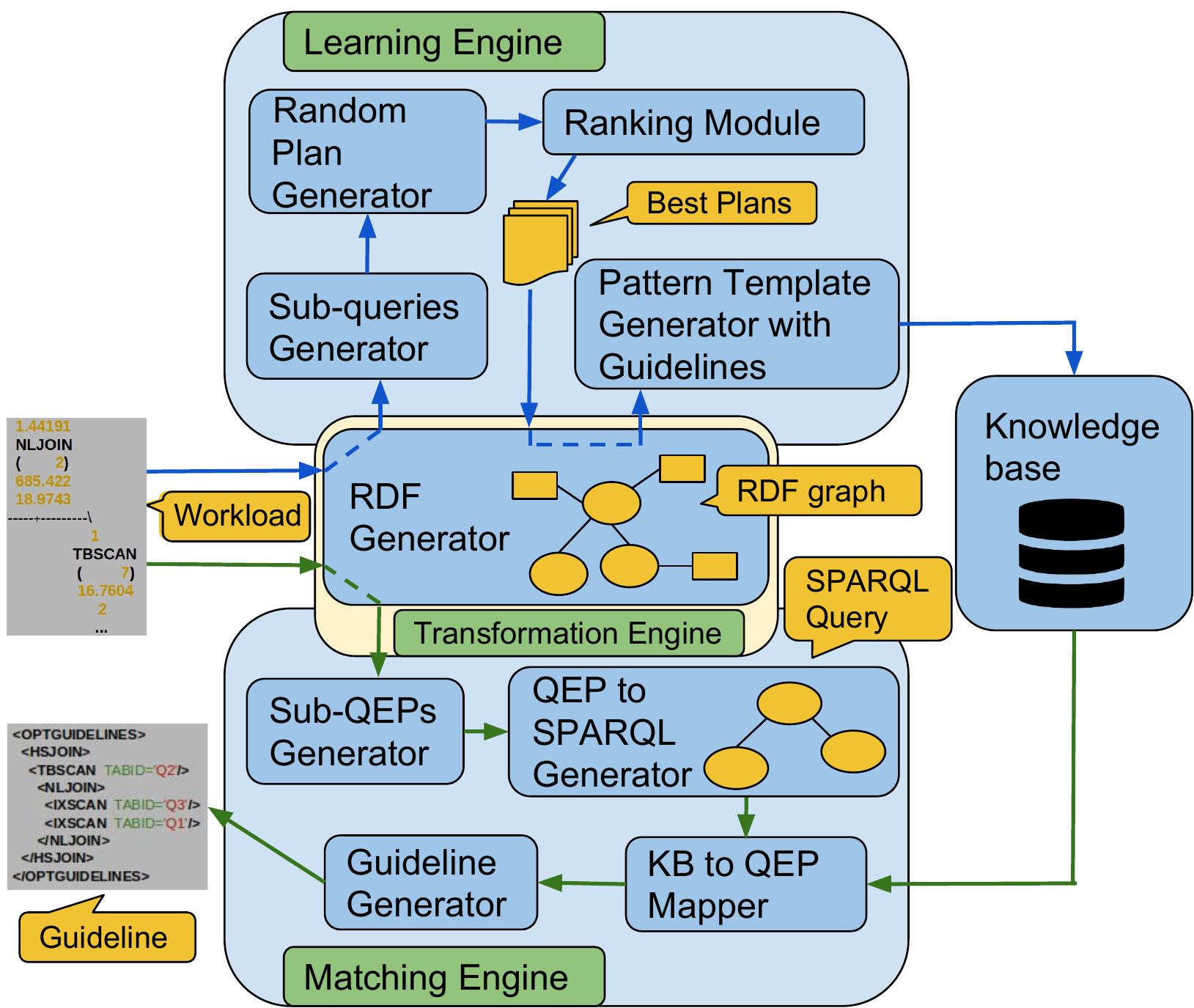}
     \caption{System architecture of GALO.}
     \label{fig:architecture}
     \paddingBottomFigure{}
 \end{figure}

\toolName{}'s system architecture is illustrated
in Figure \ref{fig:architecture}.
The system is comprised of
\begin{enumerate}[nolistsep,leftmargin=*]
\item a \emph{transformation engine},
\item a \emph{learning engine}, and
\item a \emph{matching engine}. 
\end{enumerate}
\noindent
The back-end of \toolName{} is written in Java.
The front-end is a web-based, interactive interface
written with JavaScript libraries.

The transformation engine is the primary interface
to go from SQL queries and query-execution plans
into the knowledge base and back.
The knowledge base is represented in RDF and interacted
with (\emph{queried via}) SPARQL.
The \emph{learning engine} is used \emph{offline}
to populate the knowledge base
with discovered rewrites.
The \emph{matching engine} is used \emph{online}
to match rewrites to query plans of SQL queries
from the workload queued for execution
for the purpose of re-optimization.




There are, thus, two \emph{workflows} for \toolName{}:
\emph{offline learning}
and
\emph{online re-optimization}.
In Figure \ref{fig:architecture},
\emph{offline learning} workflow
is on the top,
through the transformation
and learning engines,
for updating the knowledge base.
The \emph{online re-optimization} workflow
is on the bottom,
through the transformation
and matching engines,
employing the knowledge base.


\section{Modules} \label{section:Modules}

We detail the \emph{knowledge base}
(along with the \emph{transformation engine}),
the \emph{learning engine},
and the \emph{matching engine},
each in turn.

\subsection{Knowledge Base} \label{section:KB}

A \emph{query execution plan} (QEP)---%
or simply a \emph{query plan}, for short---%
is the executable plan constructed by the query optimizer
for an SQL query
to be evaluated at runtime.
Within IBM DB2,
query plans are represented in the \emph{query graph model} (QGM).
An SQL query is parsed into a QGM representation;
such a plan within IBM is referred to as a ``QGM''.
That QGM is then rewritten by DB2's query rewrite engine,
which applies general heuristic transformations to the ``query''
(the QGM)
known to generally simplify the query
for purposes of evaluation.
The resulting QGM is then passed to DB2's cost-based optimizer,
which annotates the QGM to a full-fledged query plan,
which constitutes a query execution plan.

A QGM can be read as a diagnostic file
as produced by the IBM DB2 optimizer.
The QGM profiles the \emph{access paths} chosen by the optimizer,
the chosen join types (e.g., sort-merge, index nested loop),
the join order,
and index usage.
The QGM is represented as a compact graph structure.
(Hence the name, \emph{query graph model}.)
Each LOLEPOP is described in detailed textual blocks identified by ID.
Figure~\ref{fig:QEP_problem_1} depicts
a portion of the QGM file that results
for a real-world SQL query from an IBM client.
Each node of the tree represents an indivisible operator (LOLEPOP),
along with its associated estimated costs.

As QGM's are essentially graphs,
representing them as such is natural.
For building, maintaining, and accessing a knowledge base
of query plans,
we want a flexible graph representation,
and a powerful, general ``API'' for accessing and maintaining
the knowledge base.
We choose then for the representation of the knowledge base, 
the Resource Description Framework (RDF).
RDF's corresponding SPARQL query language
provides the means to query and update the knowledge base.
Matching to (sub-)query plans as stored in RDF in the knowledge base
requires recursive path matching in the graph;
such \emph{regular path queries}
(called \emph{property paths} in SPARQL)
are part of SPARQL 1.1,
and are a fundamental part of the query language.

The transformation engine is \toolName{}'s general tool
to translate QGM's and SQL queries into RDF graphs
(and to translate back).
An RDF graph is (conceptually) comprised of triples:
\emph{subject} (resource);
\emph{predicate} (property or relationship);
and \emph{object} (value or resource).
As such,
an RDF triple
describes an ``edge'' in the graph
from the vertex \emph{source}
to the ``vertex'' \emph{object}
and labeled as \emph{predicate}.
	RDF also allows for the object of a triple
	to be a \emph{value}
	instead of another node.

RDF statements can describe characteristics of subjects
via predicates and values.
The relationships between LOLEPOP's of a QGM
can be thus modeled.
The entities and characteristics of the QGM are mapped
into resources, capturing the properties and relationships
between them.
The resulting RDF graph is a full transformation
of the text-based QGM.



Once modeled as an RDF graph,
SPARQL queries can be used
to find matches in the graph.
SPARQL's recursive capabilities via property paths
let us
search for loosely connected child operators
(separated by other operators),
and to match patterns that appear multiple times
throughout the same QGM.
This approach is, of course, also more efficient
than ad-hoc usage of UNIX tools, such as \emph{regex} and \emph{grep},
that experts still use to search QGM files themselves.

\toolName{} uses the Apache Jena RDF API to map
the QGM into an RDF graph.
Jena RDF API is a Java framework
that can be used for the creation and manipulation
of RDF graphs \cite{mcbride2001jena}.
Jena is a popular option in the domain.
It is an open-source framework for building linked data,
it has an API for building RDF graphs, and
it natively supports triple store servers such as Jena's Fuseki.
%
%
    \col{For example,
    consider below a portion of the RDF graph
	as translated from the QGM presented
	in Fig.~\ref{fig:QEP_problem_1_a}.
	The statement
	\begin{center}
	\textit{$<$http://galo/qep/pop/2$>$$<$http://galo/qep/property/
	has\-PopType$>$NLJOIN}
	\end{center}
	represents the subject (LOLEPOP) with ID \#2,
	the predicate \textit{hasPopType},
	and the object NLJOIN.
	This subject also contains additional properties,
	such as the estimated cardinality,
	containing the value of 2949250.
	\begin{center}
	\textit{$<$http://galo/qep/pop/2$>$$<$http://galo/qep/property/
	hasEstimateCardinality$>$ "2949250"}
	\end{center}
	This LOLEPOP connects to another LOLEPOP 
	with ID \#3 as its outer input stream.
	\begin{center}
	\textit{$<$http://galo/qep/pop/2$>$$<$http://galo/qep/property/
	hasOuterInputStream$>$$<$http://galo/qep/pop/3$>$}
	\end{center}
	%
	}

\subsection{Learning Engine} \label{subsection:learning_engine}

During the offline learning process,
\col{large and complex} SQL queries from the workload are \emph{segmented}
into sub-queries.
Random plans are generated over and benchmarked (via runtime performance)
against the optimizer's chosen plans.
When better random plans are discovered,
they are ranked to determine the best QGM for each sub-query.
Finally,
the best are abstracted into rewrites,
template patterns,
to be stored in the knowledge base.

\col{The learning engine is run \emph{offline} inside the organization,
when the resources over the systems are
not in use, or when load is low.
This includes nights and other non-peak hours, such as weekends and holidays.  We used several machines inside IBM during non-peak hours
to improve scalability by paralleling the computation.}

\begin{figure}
    \centering
    \subfloat[Sample query with joins and predicates. \label{fig:SQL_a}]{
        \includegraphics[width=0.8\linewidth]{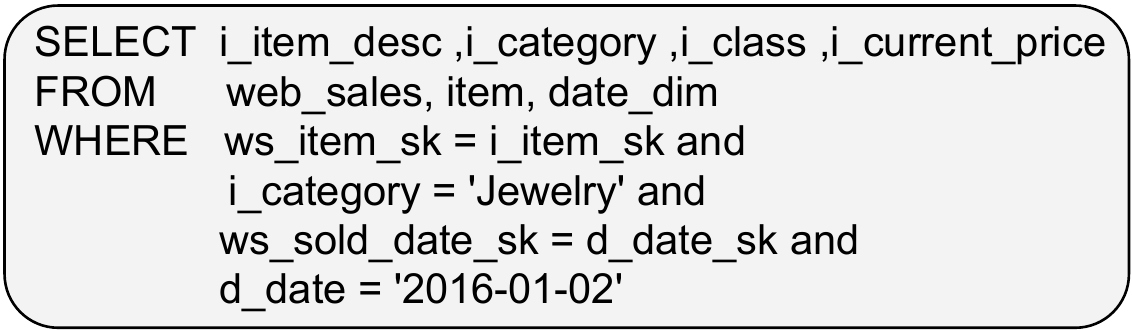}
    }
    
    \subfloat[Generated sub-query. \label{fig:SQL_b}]{
        \includegraphics[width=0.8\linewidth]{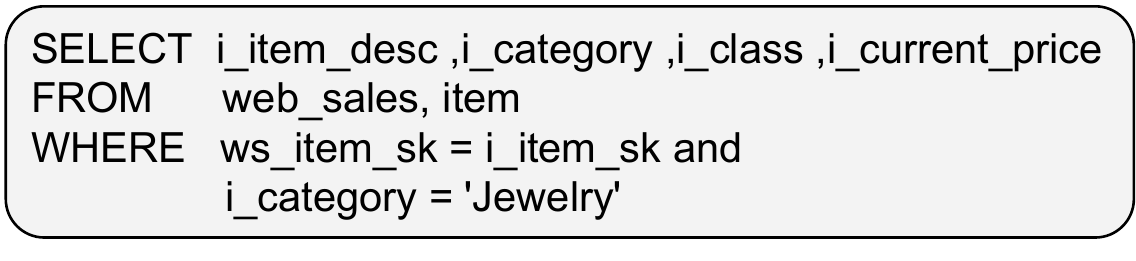}
    }
    \paddingTopFigure{}
    \caption{Sub-queries generation process.}
    \label{fig:SQL}
\end{figure}

\noindent
\textbf{Sub-query Generation.}
The learning engine is responsible for populating
the knowledge base with problem pattern templates
and their counterpart recommendations.
\col{Large} SQL queries are decomposed into \col{smaller} parts corresponding to sub-queries
to find problematic patterns
that can be applied over the query \col{workloads}
for re-optimization
(discussed in the next section).

From a given RDF-based QGM,
all SQL sub-queries are auto-generated
up to a predefined size threshold
(number of joins).
A sub-query projects the join and local predicates
from the original query that are applicable
to the sub-query's selected tables.
%
An example of the sub-query generation process
is illustrated in Figure \ref{fig:SQL}.
Figure ~\ref{fig:SQL_a} shows the original SQL query,
which consists of a three-way join between TPC-DS benchmark schema tables
\textit{web\_sales}, \textit{item} and \textit{date\_dim}.
Figure~\ref{fig:SQL_b} represents one
of the sub-queries, a two-way join
between tables \textit{web\_sales} and \textit{item}.

The system produces potential problem-pattern \emph{templates}
from sub-queries by generating over predicates' property ranges
with various cardinalities.
Property ranges are generated by sampling the database,
and are used to establish problem-pattern templates
with the same best plan
within lower and upper-bound cardinalities.
\col{This precaution is to ensure that problem patterns discovered
over one query can be used to match other queries
with different contexts of table and attribute names,
but with the same sub-structure.}
For example,
to create the property ranges
for the sub-query in Fig. \ref{fig:SQL_b},
the \textit{i\_category} attribute
in the \textit{WHERE} predicate
is sampled from the TPC-DS database
to find varying cardinalities.
For isntance, the predicate ``\textit{i\_category is NULL}'' returns
1,949 rows and ``\textit{i\_category = 'Music'}'' returns
74,426 rows. 

\col{The learning engine is designed to operate on top of dynamic data environments with changing statistics.
As data change,
the lower and upper-bound cardinalities
for problem patterns can be updated over the time
to account for cardinalities not observed before.}

\noindent
\textbf{The Ranking Process.}
For each of the sub-queries,
alternative QGM's are produced
via the Random Plan Generator
(a tool available inside IBM DB2).
Alternative plans are compared against the QGM
chosen by the optimizer as ``optimal''. As the cost estimates used during optimization are not always accurate
with respect to what is observed at runtime,
the runtime statistics are obtained by executing the alternative \col{QGMs}
via DB2's \texttt{db2batch} utility tool.
The ranking objective is to determine
the best QGM for each sub-query
within the predicate property ranges.
\col{Ultimately,
if multiple competing plans were found to be better
for a given sub-query,
the best is chosen as a rewrite
to add to the knowledge base.}

Each QGM is run multiple times to obtain an accurate baseline cost,
to remove noise related to the server or network load.
The ranking process uses K-means clustering
to remove outliers based on elapsed time.
The clustering algorithm divides QGM's into two clusters:
\emph{prospective} and \emph{anomaly}.
QGM's in the prospective cluster are then considered,
while those in the anomaly cluster are ignored.
In the case of ties,
the system considers other features as a tie breaker.
These are measures of other resource usages,
such as buffer pool data logical reads and physical reads,
total CPU time usage, and shared sort-heap high-water mark.

\begin{figure}[t]
    \subfloat[Plan obtained by the optimizer. \label{QEP_problem_2_a}]{
        \includegraphics[width=0.87\linewidth]{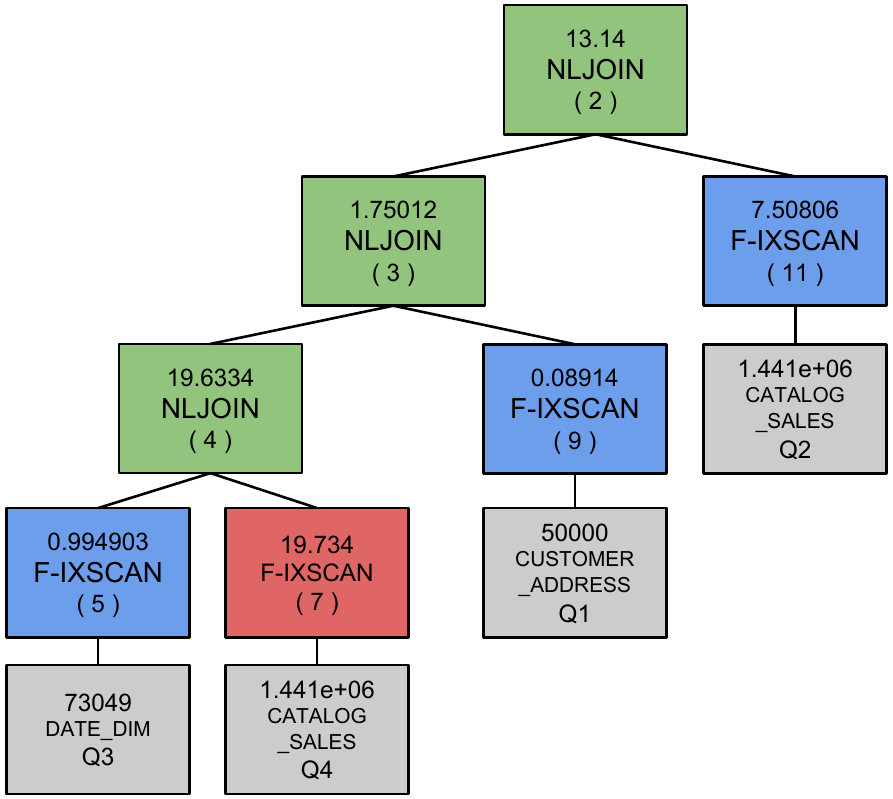}
    }
    
     \subfloat[Faster plan found by \toolName{}. \label{QEP_problem_2_b}]{
        \includegraphics[width=0.87\linewidth]{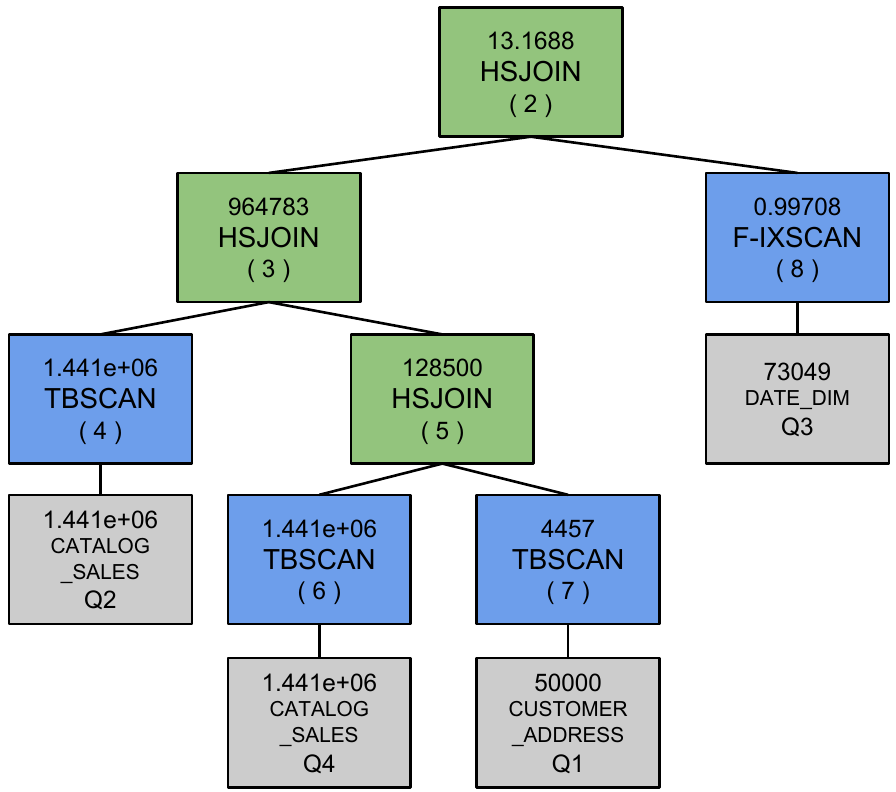}
    }
    
     \paddingTopFigure{}
    \caption{Hash-Join Bloom Filter Problem Pattern.}
    \label{QEP_problem_2}
     \paddingBottomFigure{}
\end{figure}

A problematic portion of QGM as chosen by the optimizer
from the TPC-DS workload query is shown
in Figure \ref{QEP_problem_2_a}.
At runtime,
the \texttt{F-IXSCAN} (\#7) suffers from excessive random I/O reads.
There is a \emph{flooding} problem.
This is a consequence of a poorly clustered index used
to access the \texttt{CATALOG\_SALES} table instance \texttt{Q4},
causing pages to be loaded
into the buffer pool as usual,
however, then being overwritten by other pages subsequently loaded.
When a replaced page needs to be read again,
it is subsequently loaded back into the buffer pool.
This adds significant I/O's.
This results in a poorly performing \texttt{NLJOIN} (\#4)
when joining the problematic \texttt{F-IXSCAN} (\#7)
with the \texttt{F-IXSCAN} (\#5) over the \texttt{DATE\_DIM} table instance \texttt{Q3}.
This overhead is propagated upward into the next \texttt{NLJOIN} (\#2)
and operators to follow,
causing further performance issues.

Early occurrence performance issues like this must be addressed
as they affect the whole QGM.
These are the types of challenges that experts encounter.
With a sparsity of accessible solutions,
automated methods become that much more important.
\toolName{} finds a solution to the problem pattern
in Fig. \ref{QEP_problem_2_b}.
The discovered solution applies a hash-join bloom filter
in the \texttt{HSJOIN} (\#2).
A bloom filter is a space-efficient, probabilistic data structure
to test whether an element is a member of a set
by hashing the values and performing a bit comparison between them
\cite{Barber:2014:MHJ:2735496.2735499}.
	False positives can occur;
	however, false negatives never occur.
	A bloom filter can filter so whole partitions
	never need to be read,
	as we know by a filter miss,
	nothing in the partition can match.
In the better query plan,
the  hash join creates a bitmap from the inner input.
This bitmap is used as a bloom  filter lookup for the join,
to avoid hash-table probes for outer tuples that never can match.
This results in a drastically faster execution plan.


\noindent
\textbf{Knowledge-base Generation.} 
Detected query problem patterns are transformed into \emph{templates}
to be saved in the kno\-wle\-dge-base RDF graph.
This is a critical abstraction step
that enables different queries with varying tables and predicates
later to match to patterns in the knowledge base.
\col{Table and column names are replaced
by the canonical symbol labels
in the QGM.
When SPARQL queries are generated for the matching
for online re-optimization,
the SPA\-RQL node-binding variables will match to these.
Thus, queries with the same sub-structure and characteristics,
but with different table and attribute names, are matched
against the same problem pattern template.
This assures that problem patterns usability is not limited
to a specific query or query workload.
Over time,
a \emph{unified knowledge base} of problem patterns and guidelines
learned over various query workloads is created.
Our experiments
in Section~\ref{subsection:exp:effect} (Exp--2) demonstrate
that, in practice,
problem patterns overlap significantly across various query workloads.} 
A template is generated over the predicate property ranges
with the same best plan
by sampling the database
with various cardinalities,
to establish the lower and upper bound for properties.%

The generated resources are keyed by the IDs of the LOLEPOPs
in the QGM,
different resources could potentially have name collisions
between their problem-pattern templates in the knowledge base.
To rectify this,
each resource is ano\-nymi\-zed by generating a unique random identifier.

\col{
	The upper- and lower-bound values are each stored
	in their own respective tags in the predicate.
	For instance,
	the upper-bound value
	for the \texttt{hasCardinality} property is stored
	as \texttt{hasHigherCardinality},
	and the lower-bound value is stored
	as \texttt{hasLowerCardinality}.
	Consider a portion of the RDF graph corresponding
	to the hash-join bloom filter pattern
	from Figure \ref{QEP_problem_2}.
	Our system was able to determine
	that LOLEPOP \#5
	has a cardinality lower bound of 19,771 and upper bound of 128,500.
	\begin{center}
	$<$http://galo/qep/pop/5$>$$<$http://galo/qep/property/
	hasLowerCardinality$>$ "19771"
	\textit{$<$http://galo/qep/pop/5$>$$<$http://galo/qep/property/
	hasHigherCardinality$>$ "128500"}
	\end{center}
	This problem pattern template dictates
	that any QGM that falls in the given range and matches the rest of the structure should be re-optimized.
}

The knowledge base is housed in an Apache Jena Fuseki SPARQL server.
Fuseki is a SPARQL end-point accessible via HTTP protocols.
This provides a REST (Representational State Transfer) API
for querying the knowledge-base graph on the server.
We opted for this service as it is integrated
with TDB (Native Triple Store),
a Jena component for RDF storage and querying.
While TDB can be used as a RDF storage on a single machine,
Fuseki has parallelism built in,
enabling multiple requests
to be performed concurrently.
It provides a robust, transactional, and persistent storage layer.

The recommended replacement patterns
for corresponding ``malicious'' problem pattern templates are stored
in the knowledge base as \emph{guidelines}.
A guideline document is represented as an XML document.
It imposes characteristics on the plan
during the cost-based phase of optimization,
such as the join methods (e.g., hash-join or merge-join),
join order (enforced by the order of the XML tags),
and access methods (e.g, index scan).
Join tags in the guidelines require two child elements:
the first corresponding to the outer input; and
the second to the inner input of the join.
A plan re-optimization guideline document does not
necessarily specify all aspects of the execution decisions.
Unspecified aspects of the execution plan will default
to being chosen by the optimizer in a cost-based fashion.
A guideline document can be generated by the matching engine,
then, to be provided with the SQL query
back to the optimizer for ``re-optimization''.%
\footnote{%
	Note that a guideline, in truth,
	is a strong suggestion to the optimizer.
	A guideline will not be used if 
	other previous employed guidelines
	lead to a (partial) query plan
	in which the guideline in question is no longer applicable.
}


\begin{figure}
\center
    \includegraphics[width=0.77\linewidth]{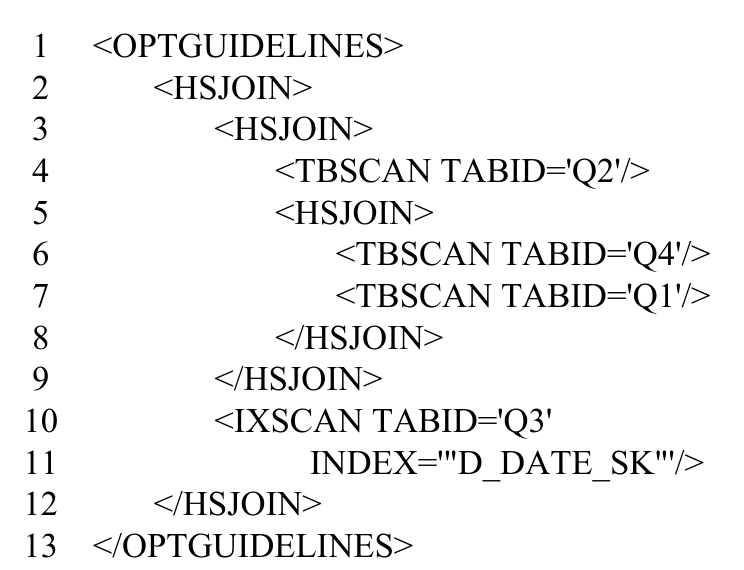}
    \paddingTopFigure{}
    \caption{Guideline generated for plan in Figure \ref{QEP_problem_2_b}.}
    \label{fig:Guideline}
    \paddingBottomFigure{}
\end{figure}

Figure \ref{fig:Guideline} illustrates the XML guideline generated
for the QGM
in Figure \ref{QEP_problem_2_b}.
The \texttt{HSJOIN} tags on Lines 2, 3 and 5, correspond
to LOLEPOP operator IDs \#2, \#3 and \#5
in Figure~\ref{QEP_problem_2_b},
respectively.
The \texttt{HSJOIN} element on Line 5 contains two child elements.
The first element indicates
that table \texttt{Q4} is an outer input of the join accessed by \texttt{TBSCAN}.
Similarly,
the second element indicates that table Q1 is
an inner input of the join,
also to be accessed with \texttt{TBSCAN}.
The \texttt{TABID} attribute specifies the table reference
to which the access should be applied.
The attribute's target table reference is
identified by its qualifier name from the QGM.
Alternatively,
the \texttt{TABLE} attribute can be used instead,
specifying the fully qualified table name.
The children of a join element does not necessarily have
to be an accessor to tables, however.
It can instead be the other join element.
The \texttt{HSJOIN} on Line 2 depicts this exact scenario.
The outer input child element (Line 3) is another \texttt{HSJOIN},
and the inner input element an \texttt{IXSCAN} (Line 10).
The latter specifies
that the optimizer should use the \texttt{D\_DATE\_SK} index
to access the \texttt{Q3} table.
(The optional \texttt{INDEX} attribute specifies the desired index
to be used in the plan).

\subsection{Matching Engine} \label{subsection:match_engine}

\noindent
\textbf{Querying Knowledge Base.} 
SPARQL is a recursive acro\-nym
for SPARQL Protocol and RDF Query Language.
As the acronym suggests,
the languages is able to retrieve data stored in the RDF format.%
\footnote{%
	SPARQL and RDF are W3C standards
	which were originally developed for \emph{semantic web}.
}
A SPARQL query consists of a set of triple patterns
similar to RDF triples.
In the query,
each of the \emph{subject}, \emph{predicate}, and \emph{object} may be
a variable.
As discussed before,
SPARQL has many language features
that are quite beneficial for \toolName{}'s tasks.
%
Since our templates are stored as RDF graphs
in the knowledge base,
querying these by SPARQL provides
an efficient way to match problematic patterns
and to retrieve the corresponding recommended patterns.



At runtime,
a \col{potentially large and complex} SQL query to be re-optimized is segmented into sub-queries
(in the similar fashion as queries are in the learning phase).
The transformation engine is used to translate the query---%
represented as an initial QGM by DB2 after the query is parsed---%
into RDF and, there, segmented.
The transformation engine then
rewrites the \col{RDF's segments} into SPARQL queries,
with the necessary characteristics to match against
the RDF problem pattern templates in the knowledge base.
(Note this is akin to a \emph{query by example}.)
The resulting SPARQL query is composed of two parts:
a SELECT clause, which defines the variables to be retrieved; and
a WHERE clause enumerating the properties to match.
Variable names in SPARQL are prefixed by ``\texttt{?}''.
To facilitate this SPARQL query generation,
we introduce handlers
to generate automatically the variable names.
We define three types of handler variables:
\emph{result},
\emph{internal}, and
\emph{relationship} handlers.

\begin{figure}[t]
\center
    \includegraphics[width=0.8\linewidth]{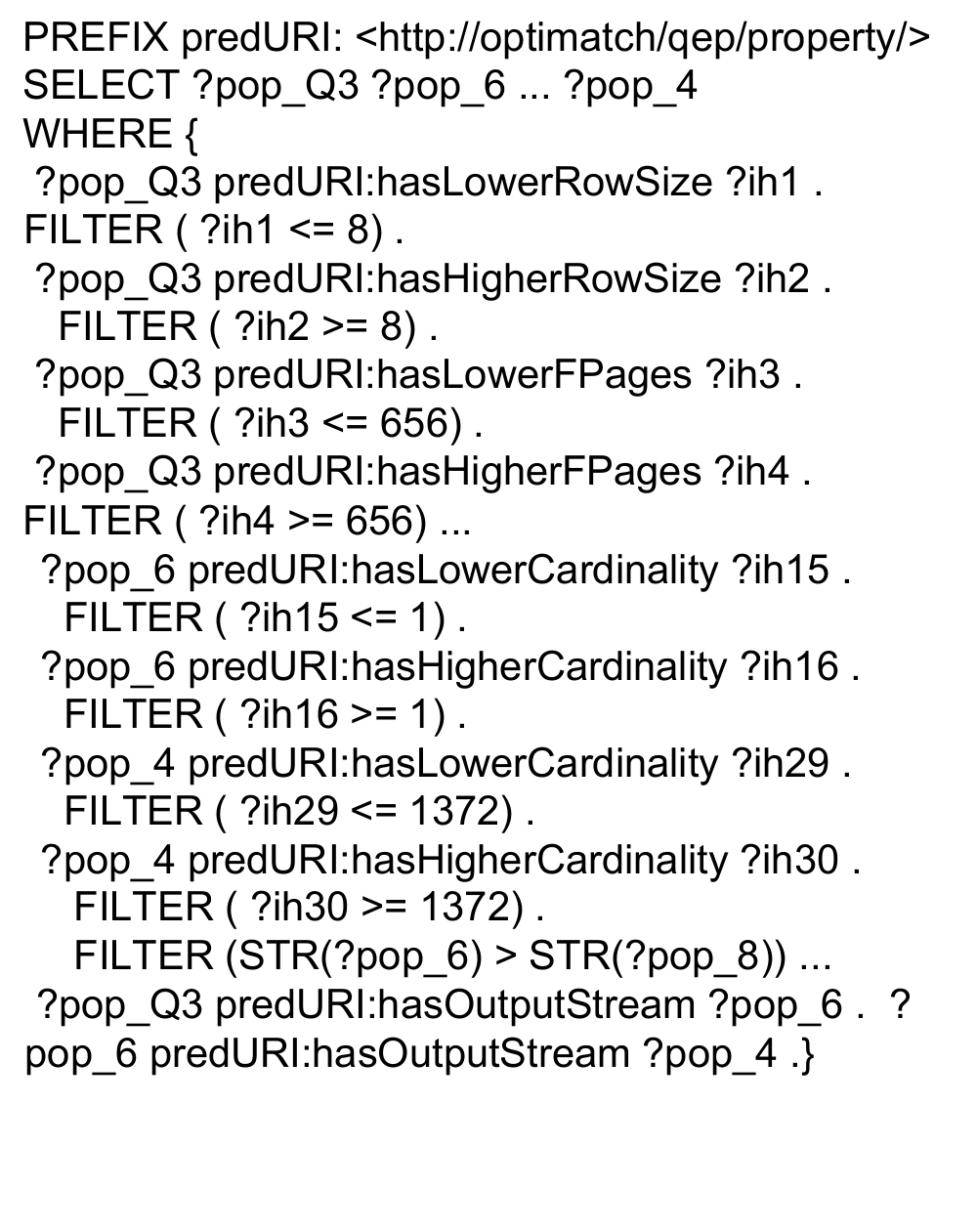}
    \paddingTopFigure{}
    \paddingTopFigure{}
    \paddingTopFigure{}
    \paddingTopFigure{}
    \caption{SPARQL query for pattern in Fig~\ref{QEP_problem_2_a}.}
    \label{Autogenerated SPARQL query}
    \paddingBottomFigure{}
\end{figure}

A \emph{result handler} names the results retrieved from the query.
It is composed of the name \texttt{pop} and the ID of the LOLEPOP
(or the name of the table instance for an access path).
This is used in the \texttt{SELECT} statement
for returning the resource,
and inside the \texttt{WHERE} statement
for creating properties of a pattern.
Figure~\ref{Autogenerated SPARQL query} illustrates
a portion of an auto-generated SPARQL query
for the problem pattern
of the QGM
from Figure~\ref{QEP_problem_2_a}.
Here,
\texttt{?pop\_Q3} and \texttt{?pop\_6} correspond
to the \texttt{TABLE} 1 table instance \texttt{Q4}
and the \texttt{IXSCAN} under the \texttt{FETCH} \#7,
respectively.
The result handler \texttt{?pop\_Q3} is a resource returned
back via \texttt{SELECT} that is also used in the \texttt{WHERE} clause
to identify the characteristics of this resource,
such as the row size
(obtained by adding the predicate \texttt{hasLowerRowSize} and
\texttt{hasHigherRowSize}).

An \emph{internal handler} is used
to aid in the stating of properties such as filtering.
We name it by ``\texttt{ih}'' (for \emph{internal handler})
appended with a sequential identifier.
In Fig.~\ref{Autogenerated SPARQL query},
an internal handler is used to filter values
for the \texttt{pop\_4} cardinalities,
first by associating it to the resource \texttt{pop\_4}
(``\texttt{?pop\_4 predURI:hasLowerCardinality ?ih29}''),
and then within a FILTER clause
(``FILTER (\texttt{?ih29} $\leq$ 1372)'').

For each property,
the SPARQL query ensures
that the values are
within the problem-pattern template range.
In Figure~\ref{Autogenerated SPARQL query},
the property \texttt{hasLowerCardinality} is used
to check the lower bound of the cardinality.
The higher bound is checked by \texttt{hasHigherCardinality},
accordingly.
The FILTER statement is used to enforce the uniqueness
of each resource via a distinct resource ID.
LOLEPOPs \#6 and \#8 are assured distinct
by applying the following filter:

\begin{center}
``\texttt{FILTER(STR(?pop\_6) > STR(?pop\_8))}''.
\end{center}

A \emph{relationship handler} establishes
a connection between nodes.
It is denoted by a \emph{result handler} in conjunction
with the property \texttt{hasOutputStream}.
In Figure~\ref{Autogenerated SPARQL query},
to connect the \texttt{FETCH IXSCAN} \#6 to the \texttt{NLJOIN} \#4
(from Figure~\ref{QEP_problem_2_a}),
the tool generates in the \texttt{WHERE} clause
a relationship statement

\begin{center}
``\texttt{?pop\_6 predURI:hasOutputStream ?pop\_4}''.
\end{center}

\noindent
\textbf{Plan Transformation.}
The QGM generated by the optimizer is
modified by matching RDF problem pattern templates.
The matches are found
by climbing up iteratively over a segmentation of the QGM
(sub-QGM's),
of the ``tree''.
The size of a sub-QGM is capped by the same predefined threshold
that was used in the learning phase (identified by the number of joins).
We verified, in practice,
that up to four joins is optimal.
This process is recursively applied
until the stopping LOLEPOP denoted as \texttt{RETURN} is found
in the QGM's RDF graph.
After all plan transformations are applied,
the recommendation guidelines are collected
into a guideline document.
The original query
coupled with the guideline document is then
passed through the optimizer again
for re-optimization.

\col{When \toolName{}\ can ``re-optimize'' a query,
it creates a guidelines document that contains the matched rewrites
from the knowledge base that apply.
This guidelines document is submitted with the query
for optimization before execution.
In this way,
the query and the guidelines are passed to the query optimizer
to produce a query execution plan.
This is a more general, and safer, way to perform re-optimization
than, say, would be adding \emph{pragma} to the query
(to represent the ``rewrites''),
which would force the chosen rewrites to be applied.
The collection of rewrites that matched might not all apply
within the plan; application of one might lead the optimizer
to an altered plan in which the others no longer apply.
This way,
the optimizer applies the guidelines that are consistent
within the course of current query planning for the query at hand.
}


\col{The \toolName{} system automates and routinizes query performance plan checks
by running a general test of all discovered problem patterns
against a given query workload.
Since problem pattern templates are abstracted
with canonical symbol names and cardinality ranges,
the system is not limited to a specific query workload,
as problem patterns learned over one query workload can be employed
in another query workload.}

\begin{figure}[t]
\center
    \subfloat[Plan selected by the optimizer. \label{QEP_problem_4_a}]{
        \includegraphics[width=0.88\linewidth]{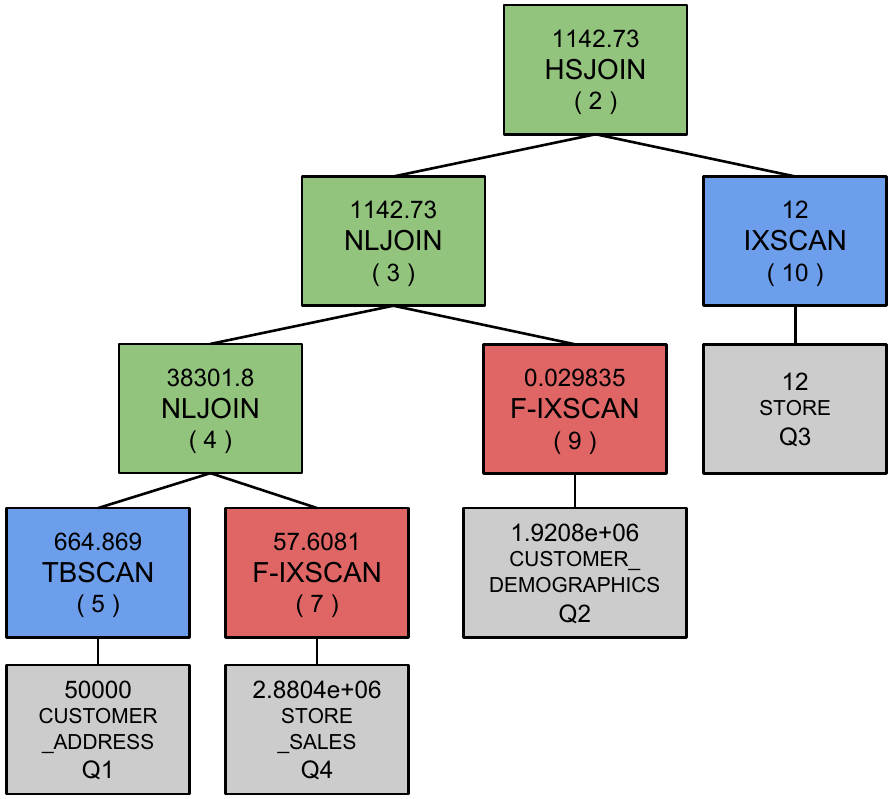}
    }
    
    \subfloat[Plan chosen by the \toolName{} system. \label{QEP_problem_4_b}]{
        \includegraphics[width=0.88\linewidth]{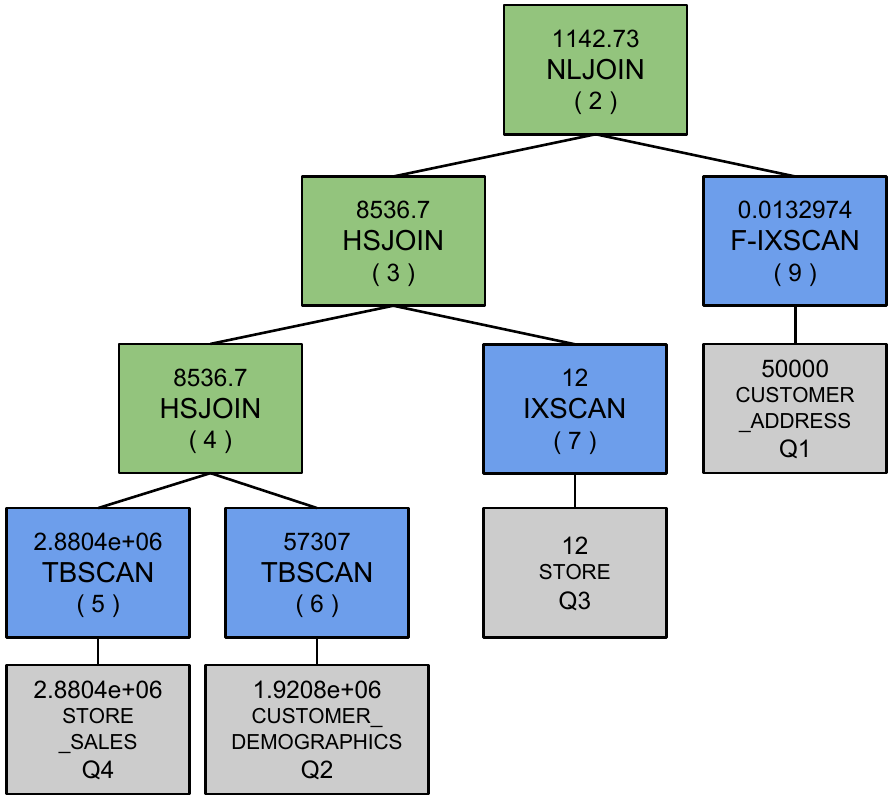}
    }
    
    \paddingTopFigure{}
    \caption{Problem pattern with transfer rate.}
    \label{QEP_problem_4}
\end{figure}

Consider the under-performing QGM as chosen by the optimizer
illustrated in Figure \ref{QEP_problem_4}
from a real-life IBM client query.
The under-performance can be mostly attributed
to a wrong estimation of the cost of the scans on tables \texttt{CUSTOMER\_DEMOGRAPHICS} (\texttt{Q2}) and \texttt{STORE\_SALES} (\texttt{Q4}).
For these tables,
the optimizer has overestimated the cost of the \texttt{TBSCAN},
and underestimated the cost of the \texttt{IXSCAN} for each table.
While favoring the \texttt{IXSCAN} would be good
for concurrently accessing data since less locking is involved
(in environments where there is large concurrency window),
compared to \texttt{TBSCAN},
we focus mainly on the actual performance of the query.
The overestimation of the \texttt{TBSCAN}
can be seen when analyzing the QGM file
represented in Figure \ref{QEP_problem_4_b}.

The \texttt{TBSCAN} \#6 over the table \texttt{CUSTOMER\_DEMOGRAPHICS} (\texttt{Q2}) has the estimate cost of 208,909,
while the total cost of the whole plan
in Figure \ref{QEP_problem_4_a}
is 207,647.
The optimizer has estimated
that the whole latter plan is less expensive
than the \texttt{TBSCAN} \#6 from the former plan.
A solution for fixing this
would be to reduce the transfer rate property in the database.
The transfer rate is a property
that refers to the transfer rate of the disk
from which the DBMS is loading the data.
Reducing the aforementioned property would fix
the overestimation of the replacement QGM
in Figure \ref{QEP_problem_4_b}.
\toolName{},
by replacing the under-performing sub-QGM
in Figure \ref{QEP_problem_4_a}
with the one
in Figure \ref{QEP_problem_4_b}.
speeds up the query execution ten times! 

\begin{figure}[t]
\center
\paddingTopFigure{}
    \subfloat[Plan obtained by the optimizer. \label{QEP_problem_3_a}]{
        \includegraphics[width=0.76\linewidth]{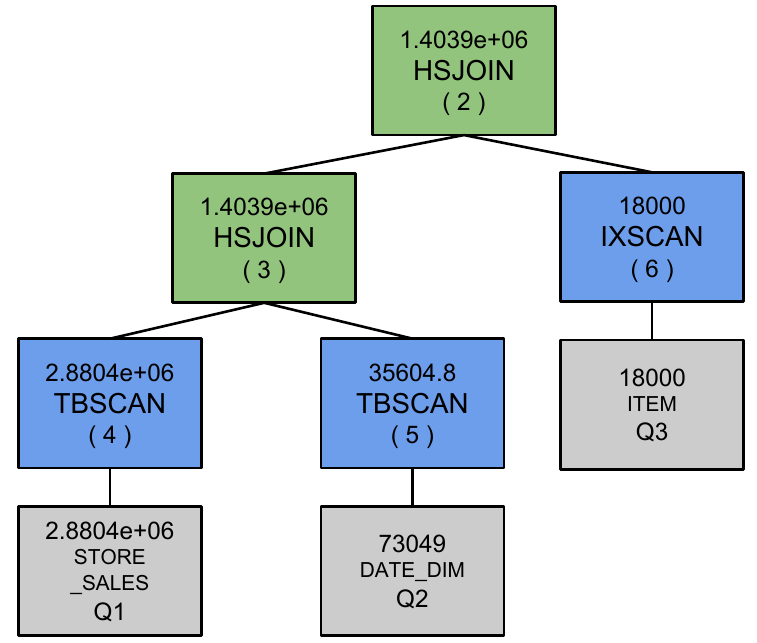}
    }
    
     \subfloat[Faster plan found by \toolName{}. \label{QEP_problem_3_b}]{
        \includegraphics[width=0.76\linewidth]{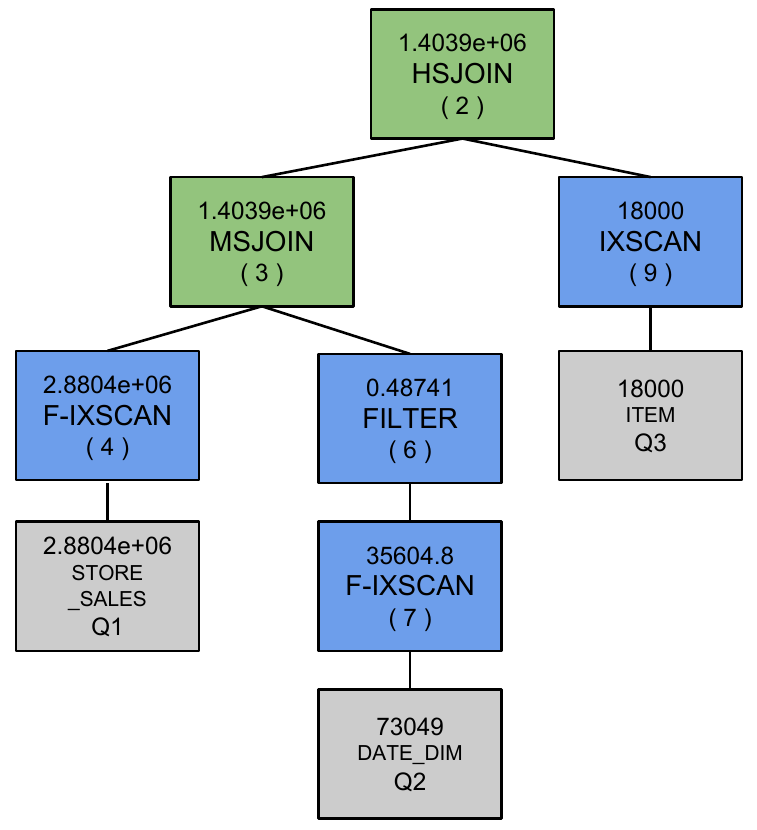}
    }
    
     \paddingTopFigure{}
    \caption{Problem pattern related to sorting.}
    \label{QEP_problem_3}
     \paddingBottomFigure{}
\end{figure}

An additional problem performance pattern related to sorting is presented in Figure ~\ref{QEP_problem_3_a} over the query from the TPC-DS benchmark. 
The problematic portion of the query is the join between the \texttt{STORE\_SALES} fact table (\texttt{Q1}) and the
\texttt{DATE\_DIM} dimension table (\texttt{Q2}). The table \texttt{DATE\_DIM} has a range of roughly
200 years and the query ranges over the first 100 years of sales. When selecting the best plan, the optimizer mistakenly assumes that 100 years of sales are matched between \texttt{STORE\_SALES} and \texttt{DATE\_DIM}, however, in practice over the instance only the last year contains sales. Thus, \texttt{HSJOIN} is chosen between \texttt{STORE\_SALES} and \texttt{DATE\_DIM} with the \texttt{TBSCAN} access method. The result is then joined 
through another \texttt{HSJOIN} with the \texttt{ITEM} table, the latter accessed via an index scan (\texttt{IXSCAN}). This execution plan suffers from the costly \texttt{HSJOIN} \#3 operation. Though the table \texttt{DATE\_DIM} is relatively small, when joined with the large table \texttt{STORE\_SALES}, 
it becomes very expensive due to the full scan on the fact table and 
the random I/Os that follow.  

A fix to this query plan found by our system is to apply \texttt{MSJOIN} between the \texttt{STORE\_SALES} and \texttt{DATE\_DIM} tables instead.
The optimization is derived from the fact that since both inputs are sorted, as soon as no more matches are found in the inner table (\texttt{DATE\_DIM}), the join operation can be safely interrupted. Our system finds this pattern as it allows to keep historical information about the estimated and actual cardinalities over operators. The scan reduction proved effective upon further
analysis over the selected plans. LOLEPOP \#4 had an initially
estimated cardinality of 2.8804+e06 in Figure \ref{QEP_problem_3_a},
which was drastically reduced to the actual 550,597 rows in
Figure~\ref{QEP_problem_3_b}, thus, providing a near 40\% overall speedup in
execution time.



\section{Experimental Study} \label{section:Experiments}


We present experimental evaluation
of \toolName{} for
    \emph{scalability},
    \emph{effectiveness},
    and
    \emph{cost} and \emph{quality}.

\begin{enumerate}[nolistsep,leftmargin=*]
\item \label{OBJ:Scalability}
    \emph{Scalability}.
    We demonstrate the scalability of \toolName{}
    with respect to varying parameters
    of the workload
    and of the knowledge base that \toolName{} builds.

\item \label{OBJ:Effectiveness}
    \emph{Effectiveness}.
    We report the performance gain
    of IBM DB2 with \toolName{}
    versus without.



\item \label{OBJ:Quality} 
    \emph{Cost and Quality}.
    We compare the rewrites learned by \toolName{}
    against those learned manually by IBM experts
    by cost of discovery and by quality of the rewrites.

\end{enumerate}

We consider the ramifications of
the workload \emph{complexity}
as measured by the number of LOLEPOP's in plans,
and the workload \emph{size}. 
We also consider the ramifications of
the knowledge-base complexity,
as measured by the number of tables to be joined
that are permitted within the segmented sub-queries. 
%
Our experiments were conducted
on servers with a 32 Intel(R) Xeon(R) CPU E5-2670
2.60GHz processor. We conducted experiments over the synthetic TPC-DS benchmark
(with 99 queries),
and real-world IBM client workload
(with 116 queries) with database size of 1GB (and main memory adjusted accordingly to simulate real-world environment). Each query was run multiple times to eliminate noise.
%




\subsection{Learning-Engine Evaluation} \label{subsection:exp:scale}

\begin{figure}[t]
\paddingTopFigure{}
    \includegraphics[width=0.94\linewidth]{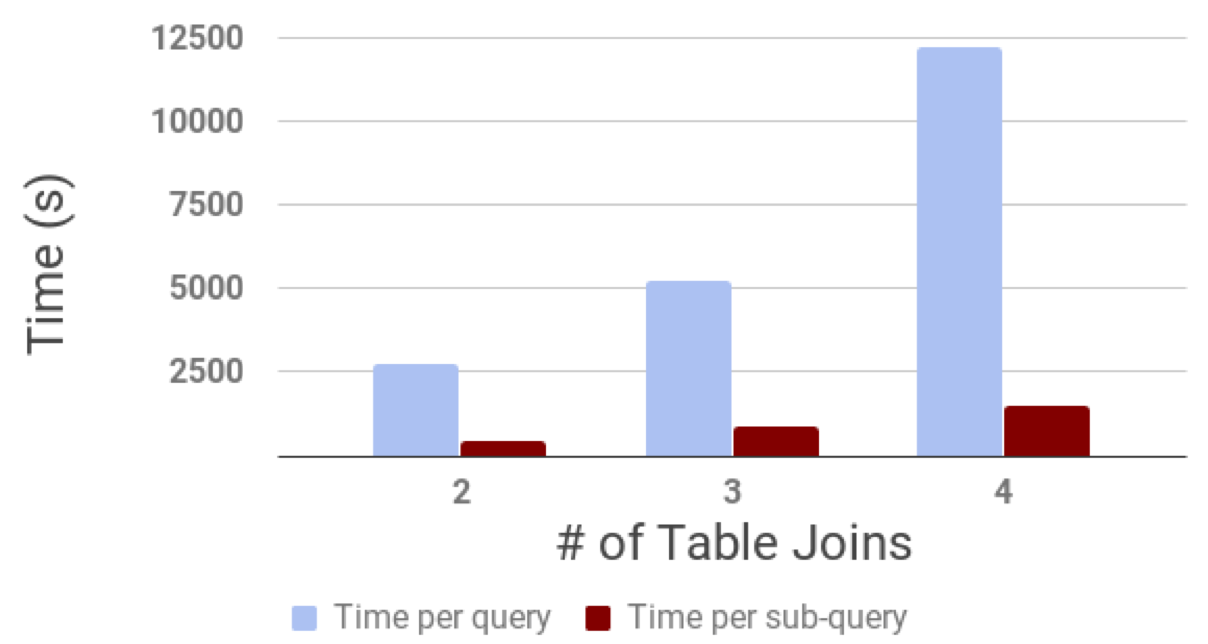}
    \paddingTopFigure{}
    \caption{Learning Scalability.}
    \label{fig:exp_n_of_tables_tpcds}
    \paddingBottomFigure{}
\end{figure}


\textbf{Exp-1: Learning Scalability and Effectiveness.}
We measure the \emph{scalability}
and \emph{effectiveness}
of the offline learning engine.
To discover rewrites,
SQL queries from the workload are decomposed
into sub-queries,
up to a predefined \emph{join-number threshold}
(number of tables to be joined).
We analyze first the results over TPC-DS,
and measure the average time to analyze each
query (and
sub-query)
with varying predicate ranges.
The analysis generates alternative plans
via the IBM DB2 Random Plan Generator.
We partition the queries
to distribute to several servers to speed up the performance. Note that the sub-queries with the same structure over different queries can be merged and evaluated once.

We report the results
in Figure \ref{fig:exp_n_of_tables_tpcds}.
%
%
The average time to analyze each query
grows exponentially as the join-number is raised
(as all combinations of joins must be considered), however, this is controlled by the table join-number threshold. The average time to analyze each sub-query grows linearly as the join-number is increased.

On the one hand,
when the join-number threshold is too low,
we do not discover pertinent problem patterns.
On the other hand,
when the join-number threshold is too high,
there are diminished returns at great expense.
Few additional problem-patterns are discovered;
and these rarely match
during the online plan re-optimization,
due to the low probability a large structure will match.
We verify that,
in practice,
a threshold of \emph{four}
provides the most optimal matching improvements.
When this threshold is held constant,
the system scales linearly
with respect to workload size and complexity.
Thus, the system scales well to large query workloads.

%



Applied to TPC-DS,
the learning engine populates the knowledge base with 98 problem pattern templates.
The average performance improvement
of the rewrites discovered for TPC-DS is 37\%.
%
We observe similar trend
over the real-world IBM client query workload
with 116 queries,
where \col{additional} 178 problem pattern templates are learned.
The average improvement
of the rewrites is 35\%.
\col{The average time per query to populate
the knowledge base is reasonable and practical,
as the computation is done offline
over the IBM systems during the non-peak hours,
and in parallel over multiple machines
(as described in Section~\ref{subsection:learning_engine})}.


 


\subsection{Matching-Engine Evaluation} \label{subsection:exp:effect}

\begin{figure}[t]
\paddingTopFigure{}
    \subfloat[TPC-DS queries.\label{fig:effectiveness_study_tpcds}]{
        \includegraphics[width=0.9\linewidth]
            {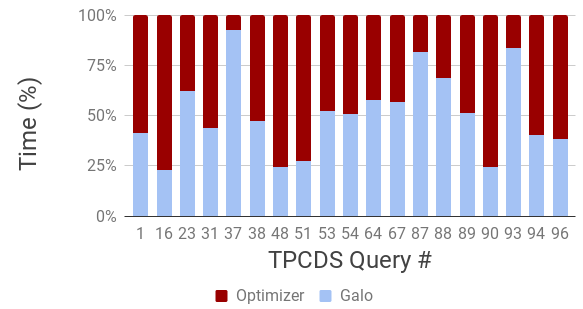}
    }
    
    \vskip -0.1cm
    
    \subfloat[IBM client queries.\label{fig:effectiveness_study_real}]{
        \includegraphics[width=0.94\linewidth]
            {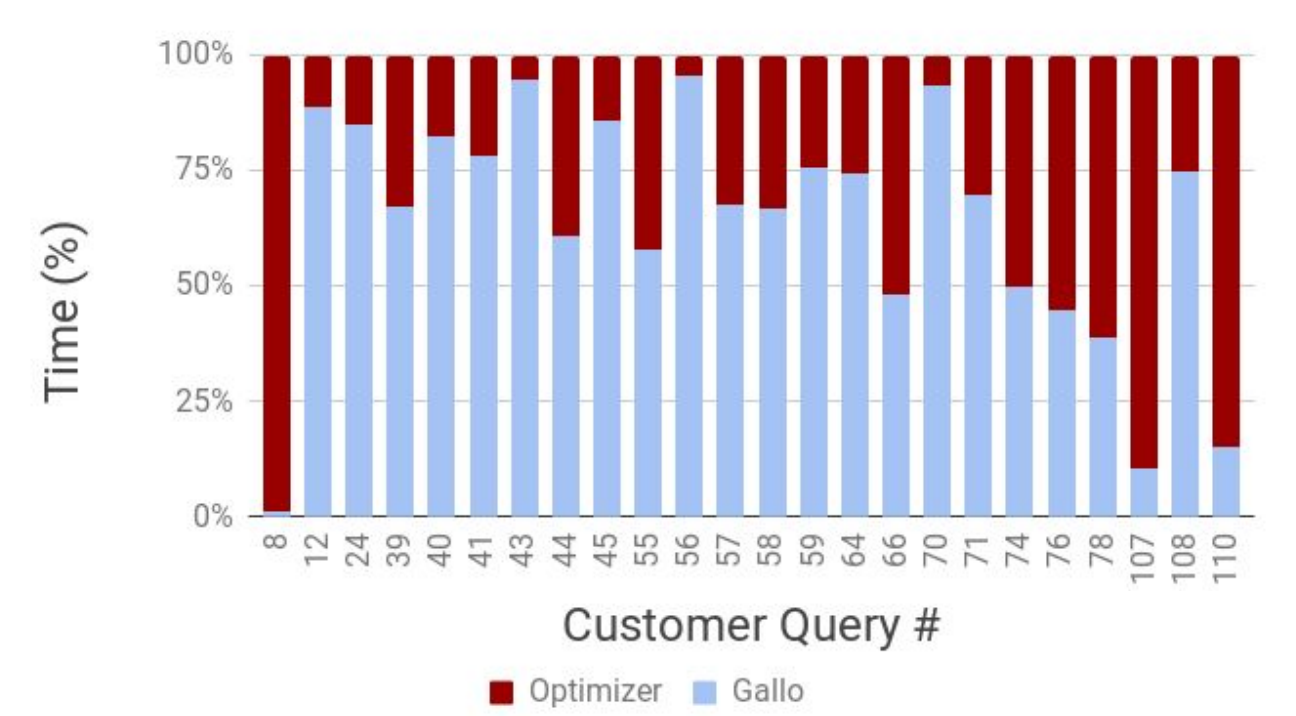}
    }
    
    \paddingTopFigure{}
    \caption{Optimizer with \toolName{} versus without.}
    \label{fig:effectiveness_study}
    \paddingBottomFigure{}
\end{figure}

\noindent \textbf{Exp-2: Matching Performance Improvement.}
We report the performance improvement
of re-optimized plans via \toolName{}
accomplished by the matching engine
compared against the plans without re-optimization
(those chosen initially by IBM DB2).
\col{We also quantify the number of problem patterns
that overlap between query workloads.}

The performance improvement results for the TPC-DS benchmark
and the real-world IBM client query workload are presented
in Figure \ref{fig:effectiveness_study_tpcds}
and in Figure \ref{fig:effectiveness_study_real},
respectively. 
The runtime for each query is normalized to ``100\%''
with respect to the runtime for the original plan.
Thus,
the red plus blue represents the runtime for the original plan,
while the blue bar represents the time
for the re-optimized plan (including the time to perform the rewrite that is marginal).


 \col{The performance gains are dramatic.
 For the TPC-DS workload the average performance gain
 by the re-optimization is 49\%.
 For a real-world IBM client workload,
 the average performance gain is 40\%.
 A significant proportion of the queries were matched for re-optimization:
 19 queries of the TPC-DS's 99 queries,
 and 24 queries of the 116 IBM client's. For instance, for query \#8 from the IBM client workload \toolName{} reduced
the query runtime from nine hours to just five minutes.}
 Performance for every one of the matched queries was improved
 by the query rewrite. 

\col{We also quantified the number of problem patterns learned
over the TPC-DS workload (Exp--1) that ma\-tch\-ed
for the re-optimization over the IBM client workload.
This experiment was performed to demonstrate
the re-usability of problem patterns learned
over different query workloads.
Interestingly,
six out of 23 queries that were improved by \toolName{}'s re-optimization
(26\%)
of the IBM client's workload
were by a rewrite
that had been learned
under the TPC-DS workload.
This validates that our system is not limited
to being workload specific.
A predetermined library of problem patterns collected
over various query workloads,
stored in the collaborative knowledge base,
can be matched against a given query workload
by adapting automatically
the dynamic context of table and attribute names. } 


\noindent \textbf{Exp-3: Matching Scalability.}
We examine the scalability
of the rewrite matching
against workloads
of varying complexity,
as measured by number of tables to join
within the workload's queries.
Queries in modern-day workloads are quite complex.
The number of tables to be joined
in the TPC-DS queries varies from one to 31.

\begin{figure}
\paddingTopFigure{}
    \includegraphics[width=0.94\linewidth]{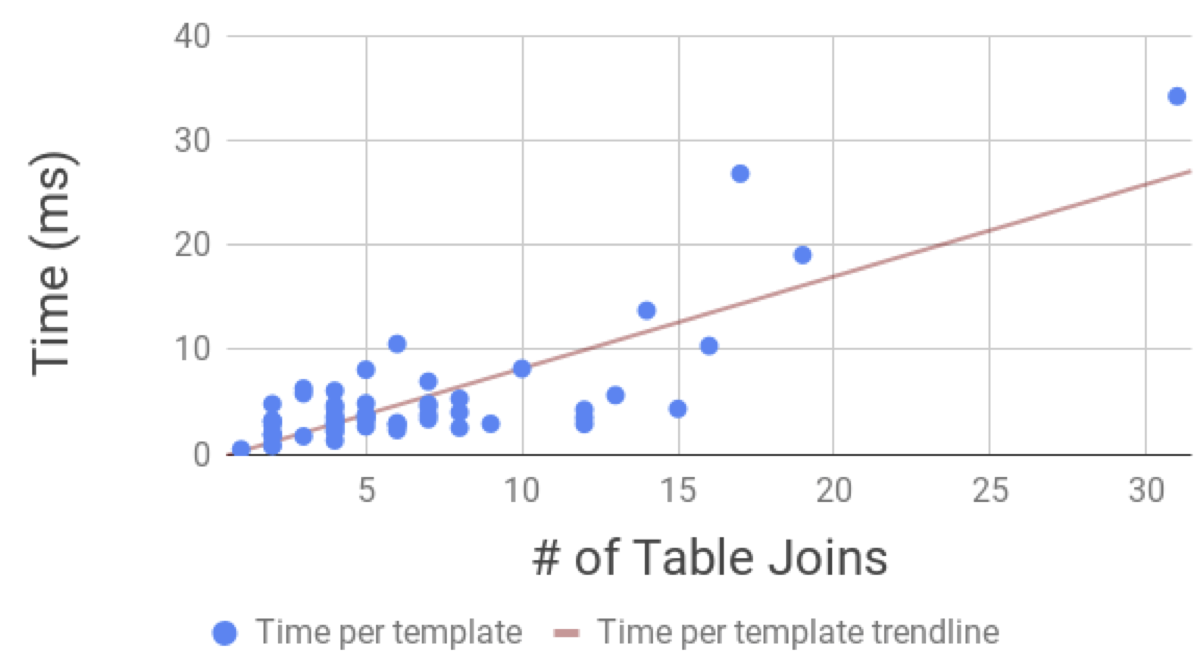}
    \paddingTopFigure{}
    \caption{Matching time in \# of table-joins.}
    \label{fig:exp_n_of_lolepops}
    \paddingBottomFigure{}
    \vskip 1mm
\end{figure}

The results are reported in Figure \ref{fig:exp_n_of_lolepops}.
The queries have been partitioned into buckets
based on their join numbers.
The average time is then reported over each bucket.
Even in the case when the queries have 32 tables to join,
the system is able to perform the match
in 34 milliseconds per rewrite.
In the less complicated case of join-number 15,
it takes 4.3 milliseconds. This cost is marginal since the time to run actual queries is minutes or hours. Overall, the trend is linear in the number joins. 
For the real-world IBM client workload, we observed similar results.

\begin{figure}[t]
    \includegraphics[width=0.94\linewidth]{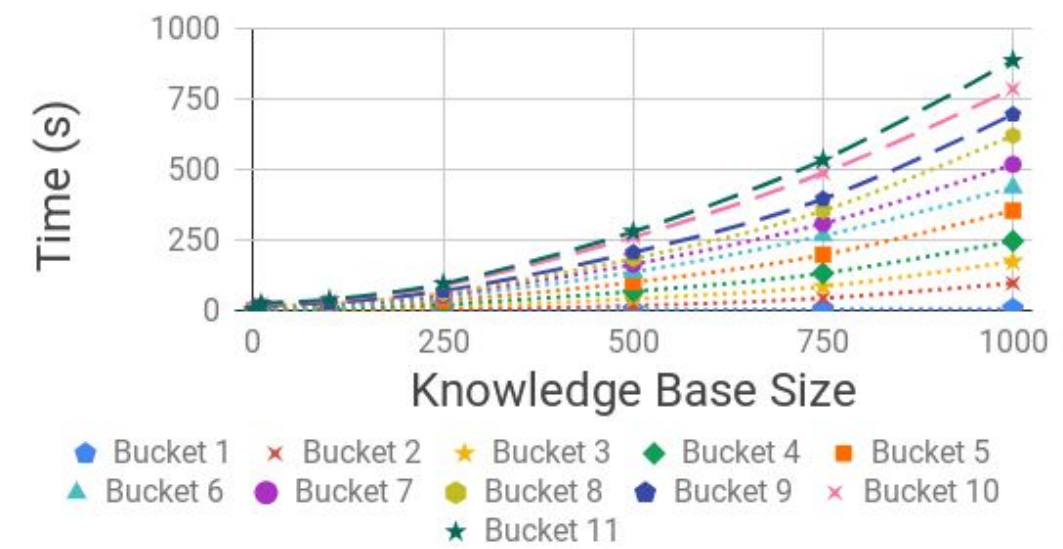}
    \paddingTopFigure{}
    \caption{Matching Engine Routinization.}
    \label{fig:exp_template_in_kb}
    \paddingBottomFigure{}
\end{figure}

\noindent \textbf{Exp-4: Routinization.}
We next examine the scalability
of the matching engine to the size of the workload
and to the number of problem patterns in the knowledge base. 
We partition the workload into buckets,
with the number of QGM's
increased by ten each time,
and the number of rewrites up to one thousand.
We report the results
in Figure \ref{fig:exp_template_in_kb}
over TPC-DS.
This shows that the system scales well
for large workloads with many problem patterns.
%
For example,
to match the 99 TPC-DS queries against the 98 learned problem patterns
in Exp-1,
the average time is 41 seconds.
For the queries of the real-world IBM client workload,
the average time to match the 116 queries
against the 178 learned problem patterns in Exp-1 is 73 seconds.
\toolName{}
can process a knowledge base
with a 1,000 problem patterns
against a workload with 100 queries
in less than 15 minutes.

\subsection{Comparative Cost \& Quality Study} \label{subsection:exp:comp}
\noindent
\textbf{Exp-5: Cost of Learning.}
We conducted a comparative study to measure
the time to perform problem determination,
both manually by IBM experts
and automatically by \toolName{}'s learning engine.
This experiment is over a sample
of four problematic queries,
due to the limited time IBM experts could spend
to participate in the experiment (as manual determination is exceedingly time consuming.)  

\begin{figure}[t]
\paddingTopFigure{}
    \includegraphics[width=0.9\linewidth]{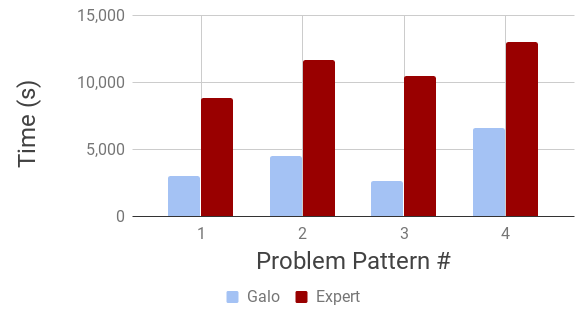}
     \vskip -2mm
    \caption{Time to learn problem patterns.}
    \label{fig:exp_comp_learn}
    \vskip -2mm
\end{figure}

We present the results
in Figure \ref{fig:exp_comp_learn}.
For the IBM experts,
we report the average time,
as four experts participated in the study.
This experiment shows that manual problem determination is
highly time consuming.
On average,
it is more than twice more expensive
than the automatic learning by \toolName{},
which can be computed offline.
Thus, using our automatic approach,
companies can save significant effort and cost as the process is fully automatic.
Note that, in many cases,
only the vendors' experts
are skilled enough
to resolve complex query performance issues. 

\begin{figure}
    \includegraphics[width=0.94\linewidth]{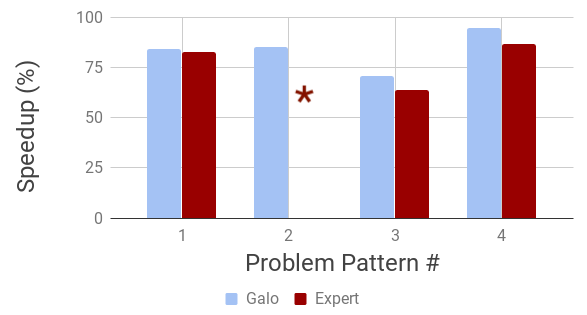}
    \paddingTopFigure{}
    \caption{Quality of learned problem patterns.}
    \label{fig:exp_comp_precision}
    \paddingBottomFigure{}
\end{figure}

\noindent
\textbf{Exp-6: Quality of Learned Problem Patterns.}
We also conducted a quality analysis
of plans obtained manually by IBM experts
and automatically by \toolName{}.
Figure \ref{fig:exp_comp_precision}
reports the percentage improvement of plans
as found manually and automatically
against the supposed optimal plans
as generated ``maliciously''
by the DB2 optimizer.

This illustrates that the manual learning is significantly
less effective than the automatic learning.
For three of the problem patterns (\#1, \#2 and \#4),
IBM experts found fixes that improved the optimizer performance;
however, the replacement plans they found are not as good
as those found by \toolName{}.
The experts were not able to find any fixes
for problem-pattern \#2 (denoted with * symbol);
\toolName{} identified and resolved the issue.  

\begin{figure}[t]
\center
    \includegraphics[width=0.85\linewidth]{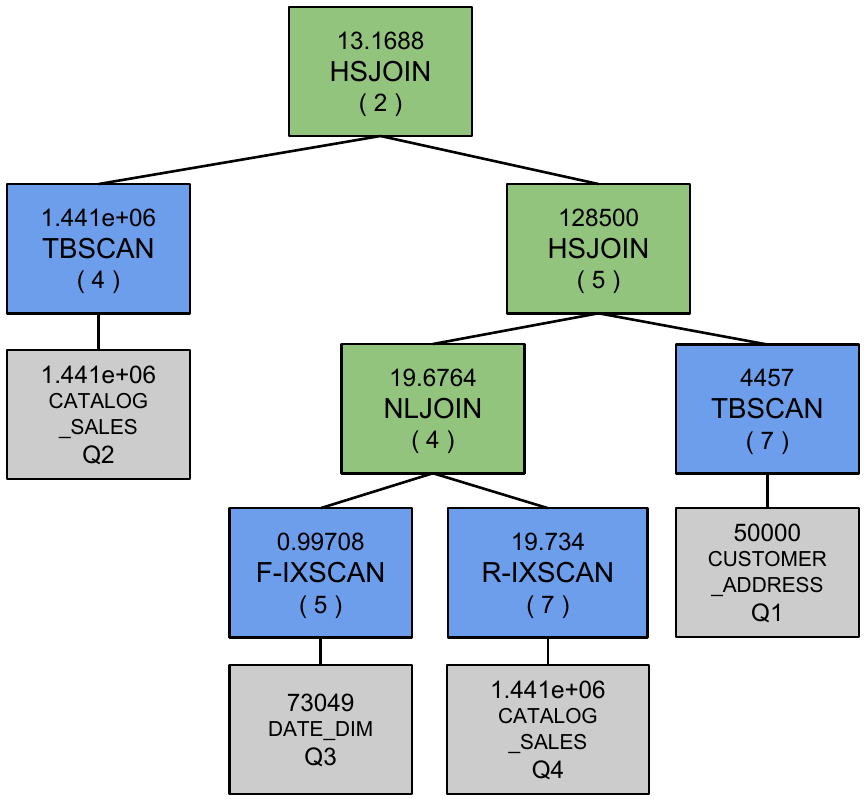}
    \caption{Expert's plan for problem in Figure \ref{QEP_problem_2_a}.}
    \label{fig:exp_comp_study_exp_qep}
    \paddingBottomFigure{}
\end{figure}

For the query
in Fig.~\ref{QEP_problem_2_a},
the experts identified the costly join in the \texttt{NLJOIN} \#2;
they changed the plan to that
in Fig. \ref{fig:exp_comp_study_exp_qep}. 
Their new plan is faster,
an 82\% improvement,
as it does not compute 
the expensive \texttt{FETCH IXSCAN}
on the \texttt{CATALOG\_SALES} table (\texttt{Q2})
for each row in the outer input.
While their improvement is significant,
the plan chosen by \toolName{} improves
over the experts' plan by another 8.6\%.

We observed also during this experiment
that problem determination is prone to human errors.
Misinterpretation was common;
for example, the value for a property in a LOLEPOP of a QGM
can be easily confused,
since it can appear in either decimal (e.g., 13.1688)
or exponential format (e.g, 1.441e+06),
as seen in Figure \ref{fig:exp_comp_study_exp_qep}.

\section{Related Work} \label{related-work}


Query optimization has been fundamental to the success
of relational database systems since their genesis.
System R
\cite{%
    selinger1979access
},
the seminal architecture for cost-based optimization,
established the importance of access-path choice, index usage,
join ordering, and pipelining,
and the use of cardinality estimation and predicate reduction factors
for guiding plan construction in a cost-based way.
Selinger's join-enumeration algoritm
at the core of System R
uses dynamic programming
to construct a query plan bottom-up,
thus ``enumerating'' through a vast plan space more efficiently.
Over the decades since,
there has been vast advances in query optimization,
both in research and development.
This work is driven
as workloads, SQL, and applications become more complex,
always moving the goal line.%

Cracks in the foundations of the mainstay approaches have begun to appear,
however,
with big data applications and ultra-complex SQL queries~\cite{DBLP:conf/ismis/GryzWQZ08,SGG+17}.%
There are two reasons for this.
First,
plans found by the optimizer are rarely now optimal.
A System-R style optimizer is guaranteed to find the best plan,
\emph{modulo} cost-estimation accuracy
    and logical compromises,
    such as not exploring bushy trees.
Estimation inaccuracy increases with complexity of queries
and system configuration.
Second,
the dynamic-programming core of the optimizer
cannot scale to very complex queries.
Our work here is directed towards addressing this first crack.

Vendors have long offered
automated tools for troubleshooting performance issues:
    IBM DB2 Design Advisor
    \cite{%
        Zilio:2004:DDA:1316689.1316783,1301362
    },
    IBM Optim Query Workload Tuner
    \cite{%
        IBM:query.Workload.Tuner
    },
    Oracle SQL Access Advisor
    \cite{%
        Dageville:2004:AST:1316689.1316784
    }, and
    Microsoft Database Engine Tuning Advisor
    \cite{%
        agrawal2005database
    }.
While such advisors are quite useful
for 
resolving general performance issues,
they generally are not fine-grained
for resolving issues at the level of plan ``debugging''.

Vendors also have introduced low-level tools for experts
to troubleshoot performance issues
for when the optimizer fails to choose optimal plans
\cite{%
    Bruno:2009:IPH:1559845.1559976,%
    Bruno:2009:4812427,%
    Dageville:2004:AST:1316689.1316784,%
    Ziauddin:2008:OPC:1454159.1454175%
}.
Oracle offers \emph{pragma} in its SQL,
and Microsoft SQL Server offers \emph{hints},
submitted with the query to override optimizer's decisions.
IBM introduced \emph{guidelines},
an XML document submitted with query to the optimizer,
to redirect the optimizer's decisions. 
However, such manual debugging is
cumbersome and time consuming,
and the performance issues are often subtle. 
The OptImatch system
\cite{%
    Damasio2016ICDE,%
    Damasio2016EDBT%
}
lets experts feed problematic query plan patterns and their resolutions
into a \emph{knowledge base}.
The knowledge base is built by hand, though.
\toolName{} automatically discovers the problem patterns. (System demonstration is described in Damasio et al.~\cite{DBC19}.)



Incorrect cardinality estimation by the optimizer is
a key factor leading to sub-optimal plans.
In \cite{DBLP:conf/cascon/LiuXYCZ15},
a neural network is applied
to improve cardinality estimation.
StatAdvisor \cite{el2009statadvisor} is a system
for recommending statistical views
for the workload and improving database statistics
that is crucial to cost-based optimizers.
Another approach to improving cost estimation is
to refine automatically the optimizer's cost model.
In
\cite{%
    Chaudhuri:1999:OQU:320248.320249,%
    He:2005:SCM:1093382.1093387
},
they introduce \emph{self-tuning} cost models.


While our work addresses the first fault of sub-optimality,
via re-optimization,
there is wide work---%
albeit primarily academic---%
on addressing the second fault of optimization scalability.
These two general efforts are orthogonal;
solutions can be combined.
A new generation of genetic algorithms
for query optimization have been introduced,
starting with
\cite{%
    Muntes-Mulero2006
},
as an alternative
to the traditional dynamic programming techniques. 
In
\cite{Marcus:2018:DRL:3211954.3211957,Ortiz:2018:LSR:3209889.3209890},
deep learning techniques are explored
for state representation and join enumeration.

\section{Conclusions} \label{conclusions}
We introduce a novel automatic system, \toolName{}, that uses RDF and SPARQL to discover problematic problem patterns and provide recommended fixes. 
\toolName{} offers a third stage of optimization, \emph{plan rewrite}.
This re-optimization leads to 
significant improvement in workload performance. 


In the future work,
we plan to apply machine learning techniques, such as deep learning,
to learn query problem patterns by varying parameters.
We also plan to develop a distributed framework
to effectively partition and load balance computation to improve further \toolName{}'s performance.




\bibliographystyle{abbrv}
\bibliography{main}  

\begin{thebibliography}{10}

\bibitem{IBM:query.Workload.Tuner}
{IBM} infosphere optim query workload tuner.
  https://www.ibm.com/support/knowledgecenter, 2018.

\bibitem{agrawal2005database}
S.~Agrawal, S.~Chaudhuri, L.~Kollar, A.~Marathe, V.~Narasayya, and M.~Syamala.
\newblock Database tuning advisor for microsoft sql server 2005.
\newblock In {\em SIGMOD}, pages 930--932, 2005.

\bibitem{Barber:2014:MHJ:2735496.2735499}
R.~Barber, G.~Lohman, I.~Pandis, V.~Raman, R.~Sidle, G.~Attaluri, N.~Chainani,
  S.~Lightstone, and D.~Sharpe.
\newblock Memory-efficient hash joins.
\newblock {\em PVLDB}, 8(4):353--364, 2014.

\bibitem{Bruno:2009:IPH:1559845.1559976}
N.~Bruno, S.~Chaudhuri, and R.~Ramamurthy.
\newblock Interactive plan hints for query optimization.
\newblock In {\em SIGMOD}, pages 1043--1046, 2009.

\bibitem{Bruno:2009:4812427}
N.~Bruno, S.~Chaudhuri, and R.~Ramamurthy.
\newblock Power hints for query optimization.
\newblock In {\em ICDE}, pages 469--480, 2009.

\bibitem{Chaudhuri:1999:OQU:320248.320249}
S.~Chaudhuri and K.~Shim.
\newblock Optimization of queries with user-defined predicates.
\newblock {\em TODS}, 24(2):177--228, 1999.

\bibitem{Dageville:2004:AST:1316689.1316784}
B.~Dageville, D.~Das, K.~Dias, K.~Yagoub, M.~Zait, and M.~Ziauddin.
\newblock Automatic sql tuning in oracle 10g.
\newblock In {\em VLDB}, pages 1098--1109, 2004.

\bibitem{DBC19}
G.~Damasio, S.~B.~V. Corvinelli, P.~Godfrey, P.~Mierzejewski, J.~Szlichta, and
  C.~Zuzarte.
\newblock {GALO}: {G}uided {A}utomated {L}earning for re-{O}ptimization.
\newblock {\em PVLDB}, 12(12):4 pages, 2019.

\bibitem{Damasio2016ICDE}
G.~Damasio, P.~Mierzejewski, J.~Szlichta, and C.~Zuzarte.
\newblock Optimatch: Semantic web system for query problem determination.
\newblock In {\em ICDE}, pages 1334--1337, 2016.

\bibitem{Damasio2016EDBT}
G.~Damasio, P.~Mierzejewski, J.~Szlichta, and C.~Zuzarte.
\newblock Query performance problem determination with knowledge base in
  semantic web system optimatch.
\newblock In {\em EDBT}, pages 515--526, 2016.

\bibitem{el2009statadvisor}
A.~El-Helw, I.~F. Ilyas, and C.~Zuzarte.
\newblock Statadvisor: Recommending statistical views.
\newblock {\em PVLDB}, 2(2):1306--1317, 2009.

\bibitem{DBLP:conf/ismis/GryzWQZ08}
J.~Gryz, Q.~Wang, X.~Qian, and C.~Zuzarte.
\newblock {SQL} queries with {CASE} expressions.
\newblock In {\em ISMIS}, pages 351--360, 2008.

\bibitem{guravannavar2005optimizing}
R.~Guravannavar, H.~S. Ramanujam, and S.~Sudarshan.
\newblock Optimizing nested queries with parameter sort orders.
\newblock In {\em VLDB}, pages 481--492, 2005.

\bibitem{He:2005:SCM:1093382.1093387}
Z.~He, B.~S. Lee, and R.~Snapp.
\newblock Self-tuning cost modeling of user-defined functions in an
  object-relational dbms.
\newblock {\em TODS}, 30(3):812--853, 2005.

\bibitem{DBLP:conf/cascon/LiuXYCZ15}
H.~Liu, M.~Xu, Z.~Yu, V.~Corvinelli, and C.~Zuzarte.
\newblock Cardinality estimation using neural networks.
\newblock In {\em CASCON}, pages 53--59, 2015.

\bibitem{Marcus:2018:DRL:3211954.3211957}
R.~Marcus and O.~Papaemmanouil.
\newblock Deep reinforcement learning for join order enumeration.
\newblock In {\em aiDM}, pages 3:1--3:4, 2018.

\bibitem{mcbride2001jena}
B.~McBride.
\newblock Jena: Implementing the rdf model and syntax specification.
\newblock In {\em ICSW}, pages 23--28, 2001.

\bibitem{Muntes-Mulero2006}
V.~Munt{\'e}s-Mulero, J.~Aguilar-Saborit, C.~Zuzarte, and J.~Larriba-Pey.
\newblock Cgo: A sound genetic optimizer for cyclic query graphs.
\newblock In {\em ICCS}, pages 156--163, 2006.

\bibitem{Ortiz:2018:LSR:3209889.3209890}
J.~Ortiz, M.~Balazinska, J.~Gehrke, and S.~S. Keerthi.
\newblock Learning state representations for query optimization with deep
  reinforcement learning.
\newblock In {\em DEEM}, pages 4:1--4:4, 2018.

\bibitem{selinger1979access}
P.~G. Selinger, M.~M. Astrahan, D.~D. Chamberlin, R.~A. Lorie, and T.~G. Price.
\newblock Access path selection in a relational database management system.
\newblock In {\em SIGMOD}, pages 23--34, 1979.

\bibitem{Si1996}
D.~Simmen, E.~Shekita, and T.~Malkemus.
\newblock Fundamental techniques for order optimization.
\newblock In {\em SIGMOD}, pages 57--67, 1996.

\bibitem{SGG+17}
J.~Szlichta, P.~Godfrey, L.~Golab, M.~Kargar, and D.~Srivastava.
\newblock {Effective and Complete Discovery of Order Dependencies via Set-based
  Axiomatization}.
\newblock {\em PVLDB}, 10(7):721--732, 2017.

\bibitem{Ziauddin:2008:OPC:1454159.1454175}
M.~Ziauddin, D.~Das, H.~Su, Y.~Zhu, and K.~Yagoub.
\newblock Optimizer plan change management: Improved stability and performance
  in oracle 11g.
\newblock {\em PVLDB}, 1(2):1346--1355, 2008.

\bibitem{Zilio:2004:DDA:1316689.1316783}
D.~C. Zilio, J.~Rao, S.~Lightstone, G.~Lohman, A.~Storm, C.~Garcia-Arellano,
  and S.~Fadden.
\newblock Db2 design advisor: Integrated automatic physical database design.
\newblock In {\em VLDB}, pages 1087--1097, 2004.

\bibitem{1301362}
D.~C. Zilio, C.~Zuzarte, S.~Lightstone, W.~Ma, G.~M. Lohman, R.~J. Cochrane,
  H.~Pirahesh, L.~Colby, J.~Gryz, E.~Alton, and G.~Valentin.
\newblock Recommending materialized views and indexes with the ibm db2 design
  advisor.
\newblock In {\em ICAC}, pages 180--187, 2004.

\end{thebibliography}
\balance

\end{document}